# Daylong sub-ambient radiative cooling with full color exterior

Suwan Jeon[1], Soomin Son[2], Seokhwan Min[1], Hyunjin Park[1], Heon Lee[2], and Jonghwa Shin[1*]

[1]*Department of Materials Science and Engineering, Korea Advanced Institute of Science and Technology, Daejeon 34141, Republic of Korea*

[2]*Department of Materials Science and Engineering, Korea University, Anam-ro 145, Seongbuk-gu, Seoul, 02841, Republic of Korea*

* qubit@kaist.ac.kr

**Abstract**

Terrestrial radiative cooling is an intriguing way to mitigate the accelerating cooling demands in the residential and commercial sectors by offering zero-energy cooling. However, the ultra-white or mirror-like appearance of radiative coolers can be visually sterile and raise safety issues when broadly applied to building facades and vehicles. To overcome the fundamental trade-off between color diversity and cooling performance, we propose a radiatively integrated, conductively insulated system that exploits thermal non-equilibrium between colorants and thermal emitters. This allows such radiative coolers to be cooled below the ambient temperature at all times of the day while exhibiting any desired exterior color including black. We experimentally demonstrate that even black coolers, absorbing 646 Wm$^{-2}$ of solar power under AM1.5 conditions, cools down to a maximum of 6.9 K (average of 3.5 K) below the ambient temperature during the daytime. These systems can potentially be used in outdoor applications, especially in commercial buildings and residential houses, where carbon-free thermal management is in high demand but diversity of colors is also important for visual appeal and comfort.

**Introduction**

In recent times, cooling demands have escalated due to economic advances and global climate change. The worst-case scenario suggests that electricity usage for space cooling will triple by 2050[1], which is alarming. Of more concern is that the majority of the cooling systems currently in use rely on electric power generated by fossil fuels, leaving large carbon footprints. To break this vicious cycle, much attention has been paid recently to radiative cooling because it makes sub-ambient cooling possible without using electricity. Its physical principle is based on both the release of radiative heat into extraterrestrial space and near-perfect solar reflection[2–9]. Comprehensive studies have examined such cooling schemes including energy-saving buildings[10,11], functional textiles[12,13], and energy harvesting systems[14,15].

In radiative cooling, however, extreme solar-reflecting conditions restrict the visual appearance of cooling surfaces to white[4–7,12], ultra-bright[16–19], or silvery[2,8] colors, which may not be suitable in practical situations due to light pollution[20], eye safety issues[21], and lack of aesthetics[22]. Several studies have explored ways to restore the color diversity in radiative cooling. A typical design is a wavelength-selective absorber design, which can suppress the solar thermal load in visible and neighboring (i.e., ultraviolet and near-infrared) wavelengths while retaining object colors[18,19,23,24]. Nevertheless, studies have shown that the diminution of solar thermal load via wavelength selection is fundamentally insufficient to realize sub-ambient cooling for the majority of colors[23,25]. Another approach is photoluminescence, which recycles the absorbed photons for color exhibition rather than converting them to heat[26–28]. In a recent study, we revealed that sub-ambient cooling is possible for all colors based on the ideal photoluminescent process[25]. While this is also a promising approach, it requires the development of optimal photoluminescent colorants for each target color. It is still not clear whether it is possible to achieve sub-ambient cooling with any color using only currently available materials.

In the current study, we present a colored radiative cooling system that can appear in arbitrary exterior colors while cooling inside objects below the ambient temperature at all times of the day, based on a simple multi-layer stacking of available materials. The different sub-systems of the cooler cooperate with one another such that, collectively, they can reject the unwanted spectral range of the sunlight, emit efficiently in the mid-infrared region while maintaining a steep temperature gradient within the system. Especially, the intentionally induced non-equilibrium state between sub-systems allows both the coloring and thermally emitting sub-systems contribute to cooling through convection and radiation, respectively. The end result is that the thermal emitter has a good radiative connection to the cold outer space but is radiatively and non-radiatively isolated from external heat sources such as the sun and atmosphere even in colored coolers. Thus, our radiatively intertwined non-equilibrium (RINE) system can resolve the conflict between color appearance and radiative cooling, in contrast to conventional radiative coolers. We experimentally show that RINE systems with dark colors (black and dark red are chosen as two examples) absorbing 646 Wm$^{-2}$ and 672 Wm$^{-2}$, respectively, under AM1.5 solar irradiation condition, accomplish sub-ambient cooling during daytime hours. Maximum temperature differences of 6.9 K and 7.4 K and average differences of 3.5 K and 3.7 K below the ambient temperature, respectively, were observed. Another notable aspect of the RINE system is that it exploits non-radiative heat exchange with the ambient air to improve its sub-ambient cooling performance, which is the opposite of conventional coolers whose sub-ambient cooling performances have been degraded by heat exchange with surrounding objects. Owing to this reversed ambient effect, the RINE system does not require additional wind covers or non-radiative shields on the exposed side, which are essential components in conventional designs to suppress non-radiative heat influx from the surrounding objects including ambient air[29].

## Results and Discussion

**Limited color diversity in current radiative surfaces for sub-ambient cooling.** We first quantify the color appearance limit of typical radiative cooling surfaces with their colorants nearly in thermal equilibrium with their thermally emissive parts for sub-ambient cooling. The temperature of radiative coolers in an outdoor environment is shown in Fig. 1a and is determined by heat exchange with primarily the sun, atmosphere, and outer space. For simplicity, we assume that the other environmental objects are at the ambient temperature ($T_{amb}$), so their effect can be included in that of the atmosphere. Among the possible pair-wise interactions, the thermal emission of the coolers toward the cold outer space is the key process in radiative cooling. However, its maximum power density $W_{max,Tamb}$ at $T_{amb}$ (which is 109 Wm$^{-2}$ at 300 K) is typically an order of magnitude smaller than the solar irradiance under AM1.5 (1000 Wm$^{-2}$) or mid-latitude summer at noon (1068 Wm$^{-2}$ for horizontal surfaces) conditions. When the radiative cooler is at $T_{amb}$, the heat exchange with the atmosphere can be ignored. Thus, for sub-ambient cooling to occur, the solar absorption of the radiative surface (white arrows in Fig. 1a) should be suppressed below $W_{max,Tamb}$ (black arrows in Fig. 1a), which means that ~10% only of solar irradiance can be absorbed by the cooler. However, this conflicts with diversifying the colors of the coolers because dark colors require a significant amount of incident light to be absorbed, far exceeding the $W_{max,Tamb}$ limit[23,25]. Consequently, conventional radiative cooler design for sub-ambient cooling has to have a mirror-like appearance or if diffusive, be pure white, or some other very bright color, absorbing less than $W_{max,Tamb}$. For instance, a radiative cooler with a bright pink color (left in Fig. 1a) can be cooled below a $T_{amb}$ of 300 K if its spectral absorptance is optimally designed to have a solar thermal load of 53 Wm$^{-2}$, which is less than $W_{max,300}$ (Fig. 1b). On the other hand, the dark reddish radiative cooler (middle in Fig. 1a) never reaches sub-ambient temperature at noon owing to a large solar thermal load of 191 Wm$^{-2}$ over $W_{max,300}$ even with the optimal spectral absorptance that achieves the lower bound of the solar thermal load of that color by the metamer optimization algorithm (Supplementary Note 1). Investigation of the sub-ambient cooling possibility of all colors in the absorptive color space reveals that the color variety in the radiative surface horizontally placed on the ground is fundamentally limited to 5% of the total color space volume, if designed to achieve sub-ambient cooling at noon in mid-latitude summer (or 14% under AM1.5 conditions as described in Ref. 25) (Supplementary Note 1).

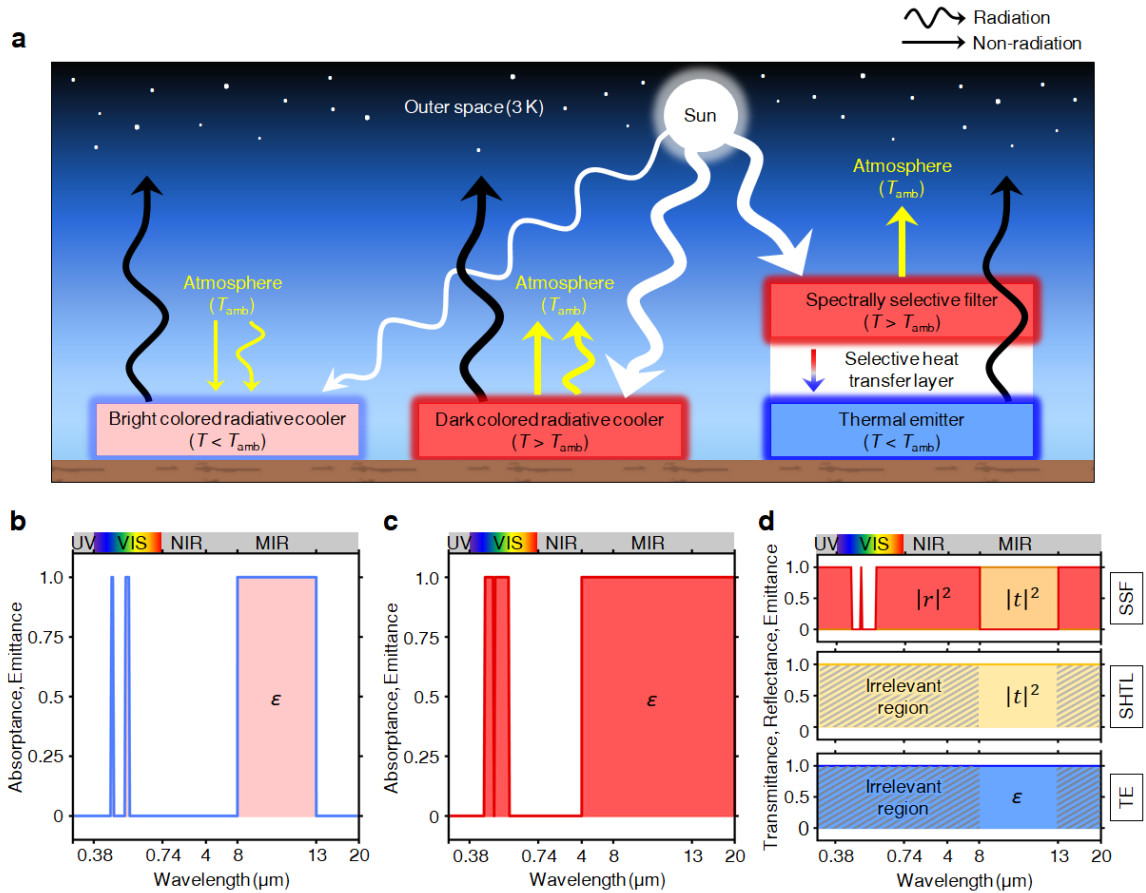

**Fig. 1 Schematic of colored radiative coolers. a** Radiative and non-radiative heat exchange scheme with the sun (white arrows), atmosphere (yellow arrows), and outer space (black arrows). **b−d** Spectral profiles for: colored selective emitter (**b**), colored broadband emitter (**c**), and RINE system (**d**). The spectrum in (**b**) and (**c**) are optimized for minimum solar absorption and maximum thermal emission. The spectrum of the SSF below 4 μm in (**d**) is matched with that of (**c**). The irrelevant regions in (**d**) corresponds to zero transmittance of the SSF where the SHTL and TE are radiatively isolated from the outside.

**RINE system.** We introduce a RINE system that exploits the thermal non-equilibrium between its color and mid-infrared emission parts for full-color sub-ambient cooling. The system comprises an outer-most spectrally selective filter (SSF), a selective heat transfer layer (SHTL) in the middle, and a thermal emitter (TE) at the back in direct contact with the objects to be cooled as shown in Fig. 1a. On the exposed side, the SSF displays colors by wavelength-selectively reflecting parts of the incident visible light and absorbing other parts, with zero transmittance at all visible wavelengths (Fig. 1d). For invisible wavelengths, the SSF functions as a selective radiative window, transmitting light in the atmospheric transparency window (typically, 8–13 μm in wavelength) and reflecting the other wavelengths. With this spectral filtering, the TE can release radiative heat toward the outer space while being radiatively insulated from other heat sources such as the sun and the atmosphere. This design approach allows the use of common broadband thermal emitters such as most polymers or water as TE materials. These materials need to have high emittance in the transparency window but their emittance at the wavelengths in the grayed-out, irrelevant region of Fig. 1d do not affect the thermodynamic state of the system. Alternatively, the design of the SSF can be simplified if the role of wavelength-selective reflection of unwanted spectral ranges of visible and infrared light is delegated to the TE or the SHTL, at the expense of increased complexity in the material selection and structural design of the TE or SHTL (Supplementary Note 2).

The key component in maintaining thermal non-equilibrium is the SHTL subsystem in the middle. This suppresses heat exchange between the hot SSF and cold TE regions while allowing thermal radiation to pass through. While the SSF and TE may each have locally near-uniform temperature distribution, the SHTL can exhibit a steep temperature gradient owing to the potentially large temperature difference between the SSF and the TE. Hence, the constituent materials for the SHTL should have as large a thermal resistance as possible to minimize conductive heat flux while being transparent to thermal infrared light.

To analyze this non-isothermal system, we establish coupled equations for net departing power densities of the SSF ($P_{SSF}$) and TE ($P_{TE}$) (both in units of $Wm^{-2}$), which consider radiative and non-radiative interactions between subsystems and with outside environments, as

$$P_{SSF} = W_{SSF} + \phi_{SSF,atm} + \phi_{SSF,SHTL} \quad (1)$$
$$P_{TE} = W_{TE} + \phi_{TE,SHTL} + \phi_{TE,inside} \quad (2)$$

where $W_{SSF}$ and $W_{TE}$ are radiative contributions to the net departing power densities from the SSF and TE, respectively, and the non-radiative heat flux densities ($\phi_{SSF,atm}$, $\phi_{SSF,SHTL}$, $\phi_{TE,SHTL}$, and $\phi_{TE,inside}$) are also defined in the outgoing direction from the subsystem of interest (either the SSF or TE) to the adjacent object identified in the subscript (detailed formulations can be found in the Methods section). The steady-state temperatures of the SSF and TE can be determined from $P_{SSF} = P_{TE} = 0$ conditions. We note that this non-isothermal description reverts to the simpler descriptions in previously reported isothermal systems if the thermal resistance between the SSF and TE is negligible such that the temperatures of the TE and SSF are effectively the same (Supplementary Note 3).

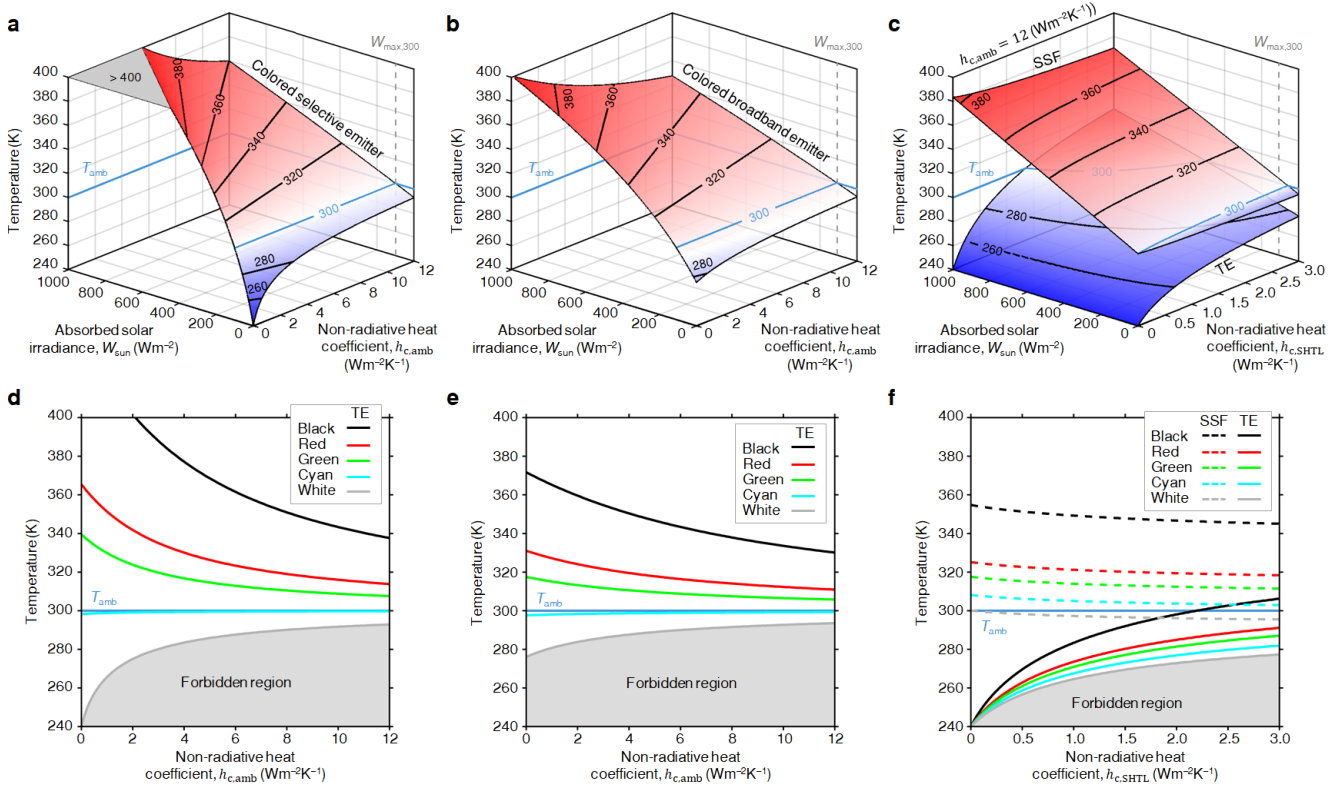

**Fig. 1 Steady-state temperature of colored radiative coolers. a–c** Steady-state temperature depending on the absorbed solar irradiance ($W_{sun}$) and non-radiative heat coefficients ($h_{c,amb}$ and $h_{c,SHTL}$) for: colored selective emitter (**a**), colored broadband emitter (**b**), and RINE system (**c**). $W_{sun}$ in (**c**) is for the SSF (or RINE system) rather than for the TE. **d–f** The lowest temperature to generate target colors (black, red, green, cyan, and white) where $W_{sun}$ accords with the lower bound of each color for: colored selective emitter (**d**), colored broadband emitter (**e**), and RINE system (**f**). The forbidden regions in (**d–f**) are inaccessible areas of each design that follow passive and reciprocal heat exchanges. In (**c**) and (**f**), $h_{c,amb}$ is fixed at 12 Wm$^{-2}$K$^{-1}$.

**Analysis for the steady-state temperature.** We evaluated and compared the cooling performance of the three colored radiative cooling schemes, as illustrated in Fig. 1. Given their spectral emittance in the mid-IR range, their steady-state temperatures can be determined by solving the above coupled equations for a given solar thermal load ($W_{sun}$) and non-radiative heat coefficients ($h_{c,amb}$ for the ambient and $h_{c,SHTL}$ for the SHTL). We assume that the TE is in thermal equilibrium with the inside object to be cooled (i.e., $\phi_{TE,inside} = 0$). Figure 2a shows the steady-state temperature of an isothermal selective emitter as a function of $W_{sun}$ and $h_{c,amb}$, with unity emittance in the 8–13 μm window and zero emittance at other mid-IR wavelengths. As $W_{sun}$ increases from zero, the temperature of the selective emitter monotonically rises and exceeds the ambient temperature ($T_{amb}$ = 300 K) when $W_{sun}$ surpasses 101 Wm$^{-2}$. Increasing $h_{c,amb}$ to strengthen the non-radiative heat exchange with the environment lowers the absolute temperature difference between the cooler and the environment. However, the control of $h_{c,amb}$ does not change the fact that the cooler temperature is above or below the ambient temperature for a given $W_{sun}$. The colored broadband emitter with a unity emittance at all wavelengths over 4 μm can be cooled more (less) than the selective emitter when $W_{sun}$ is higher (lower) than 95 Wm$^{-2}$, as shown in Fig. 2b. This confirms the intuitive understanding that broadband emitters are better at above-ambient cooling conditions and selective emitters for below-ambient cooling. However, as with the selective emitter, whether the cooler is cooled below or heated above the ambient temperature is not affected by $h_{c,amb}$, and the cooler is in fact heated, not cooled, for most parts of the possible range of $W_{sun}$ in both designs.

In Fig. 2c, we illustrate the steady-state temperatures of the SSF and TE in the RINE system as a function of $h_{c,SHTL}$, and $W_{sun}$, with $h_{c,amb}$ = 12 Wm$^{-2}$K$^{-1}$. The results reveal two critical differences compared to previous isothermal emitters. First, $W_{sun}$ is no longer limited to ≲ 100 Wm$^{-2}$ for sub-ambient cooling. Second, better insulation of the emitter from the environment or other subsystems of the cooler (lower $h_{c,amb}$ for isothermal emitters and lower $h_{c,SHTL}$ for the RINE system) always results in a lower temperature of the emitter for all considered $W_{sun}$. Both of these differences are the result of the non-equilibrium nature of the cooling system. While the SSF may be heated above the ambient temperature, the TE can be cooled below the ambient temperature if $h_{c,SHTL}$ is small. More specifically, with $h_{c,SHTL}$ < 0.72 Wm$^{-2}$K$^{-1}$, the TE reaches sub-ambient temperatures even if the SSF absorbs almost all the sunlight with $W_{sun}$ = 1000 Wm$^{-2}$, which is nearly ten times as high as the isothermal emitters' solar absorption limit for sub-ambient cooling. Because the TE temperature is always lower than the ambient temperature, a better insulation further reduces the temperature of the cooled object or results in more cooling power, regardless of the sun condition. For isothermal emitters, better insulation results in lower cooling temperatures or more cooling power only for small $W_{sun}$ values ($\lesssim 100$ Wm$^{-2}$). This makes the insulation design difficult because the requirement is the opposite at nighttime (small $h_{c,amb}$ is desirable for $W_{sun} \lesssim 100$ Wm$^{-2}$) compared to daytime (large $h_{c,amb}$ is desirable for $W_{sun} \gtrsim 100$ Wm$^{-2}$) for most colors. We also note that the broadband TE in the proposed design can reach a steady-state temperature that is lower than the lower bound of the steady-state temperature of the conventional broadband emitter in Fig. 2b, even if they have the same broadband emittance in the mid-IR range. This is due to the spectral filtering properties of the SSF, which allows the use of a simple, blackbody-like broadband TE instead of an optimal selective emitter with a precisely designed spectral emittance.

As illustrative examples, we present the expected temperatures of each scheme for five representative colors assuming optimized solar absorption spectra (i.e., $W_{sun}$ is set to its lowest possible value for each given color) (Supplementary Note 1). Figures 2d and 2e show that chromatic exhibition of black ($L^* = 0$, $a^* = 0$, $b^* = 0$ in the CIELAB color space), red ($L^* = 55$, $a^* = 80$, $b^* = 68$), and green ($L^* = 87$, $a^* = -88$, $b^* = -109$) colors overheats both selective and broadband isothermal emitters because $W_{sun}$ is much greater than 100 Wm$^{-2}$ for these colors. The steady-state temperature of cyan color is very close to $T_{amb}$, and, among the five color choices, only the white-colored ($L^* = 100$, $a^* = 0$, $b^* = 0$) isothermal emitter exhibited appreciable sub-ambient cooling performance. As can be seen from these examples, the upper bound of the solar absorption in isothermal emitters under sub-ambient cooling constraints fundamentally limits their color diversity, allowing only white or very pale colors, regardless of the level of thermal insulation. On the other hand, our RINE system can express all colors in the absorptive color space, including pure black, and achieve sub-ambient cooling at the same time if $h_{c,SHTL}$ is below 2.36 Wm$^{-2}$K$^{-1}$, as shown in Fig. 2f.

The advantage of the non-equilibrium configuration of the RINE system becomes more intuitively understandable if we look at the direction of the non-radiative heat transfer between the colorant and the ambient. Under sub-ambient cooling conditions, the colorants in the isothermal emitters had a lower temperature than the ambient. Therefore, the direction of non-radiative heat exchange with the ambient is towards the colorants and adds to the thermal load already present on the emitter-colorant composite system because of the solar absorption required for coloration. In contrast, in most situations, except for very pale colors or dim lighting conditions, the colorants in the SSF have a higher temperature than $T_{amb}$ because of solar absorption. Consequently, non-radiative heat transfer helps reduce the thermal load on the system. Thus, a large $h_{c,amb}$ is beneficial as it can lower the temperature of the SSF by dissipating heat into the ambient environment. With a high $h_{c,amb}$, the temperature difference between the SSF and the TE is reduced, which diminishes the non-radiative heat flow from the SSF to the TE, resulting in improved cooling performance of the TE (Supplementary Note 4). In contrast, the previous approach to achieve good sub-ambient cooling performance was to decrease $h_{c,amb}$ as much as possible by implementing non-radiative shields, such as air encapsulated by a thin polymeric sheet[2,4,5,7,8,14] and aerogel[6,30], to the isothermal emitters. Although our design requires good internal insulation between the SSF and the TE, it does not require special thermal insulators on the exterior, as found in previous studies[29]. This can be beneficial in practical applications that have other mechanical and chemical requirements such as scratch resistance of the exterior surface or reliability in harsh environments.

**Experimental design.** Following the theoretical studies above, we designed an exemplary non-equilibrium radiative cooling system with dark colors. The system comprises three vertically integrated sub-systems and is uniform (the SSF and TE) or periodic (the SHTL) in horizontal directions as shown in Fig. 3a. First, for the SSF design, germanium was used as a mid-IR transparent, visibly absorbing rigid substrate. To express the target color and enhance the transmission of 8–13 μm wavelength mid-IR light, we applied five stacked layers of ZnS and Ge using magnetron sputtering on the exterior surface of the substrate. The thicknesses were chosen based on needle optimization targeted at black and dark red colors (as explained further in the Methods section). We also deposited a single 1100 nm thick ZnS layer on the opposite side of the substrate for anti-reflection of mid-IR light in the atmospheric transparency window. As with commercially available multi-layer dielectric filters, more layers of coating on both sides of the substrate would result in improved solar reflection as well as better mid-IR transmittance characteristics. However, we limited them to five and one layers to demonstrate that the RINE principle can have a significant effect even with simple, readily realizable structural designs. A visual inspection in Fig. 3b of the fabricated samples revealed that they exhibited the targeted colors, and their cross-sectional images captured by scanning electron microscopy in Fig. 3c confirmed the layer configurations as designed. To check the spectral properties in the solar and mid-IR ranges quantitatively, we measured the SSFs with UV-NIR and FT-IR spectrometers, respectively; incident angles for reflectance and transmittance were set at 30° and 0° from the normal axis of the sample, respectively. In the solar spectral range, Fig. 3d shows two SSFs presenting distinctive spectral reflectance, and both have almost zero transmittance below 1.6 μm wavelength. Based on the measurement results, the absorbed solar power densities can be calculated under AM1.5 conditions for SSF-black and SSF-red, which are 646 Wm$^{-2}$ and 672 Wm$^{-2}$, respectively. This is more than six times the maximum allowed value of $W_{sun}$ for isothermal emitters for sub-ambient cooling. In the mid-IR region, Fig. 3e shows that both SSFs exhibit high transmittance in the 8–13 μm atmospheric transparency window and low transmittance at other wavelengths.

To implement the SHTL with high thermal resistance in the vertical direction, we adopted an air-and-frame approach, in which the frame comprised expanded polystyrene insulation (EPS) walls in a square-grid pattern, and the remaining volume was filled with air. While we used air at atmospheric pressure, reducing the pressure could further increase the thermal resistance. We covered the side surfaces of the EPS frame with thin Al foil (thickness ~18 μm) to reflect the radiative emission of the TE at high angles towards the outside. In addition, the Al foil was sectioned into a few horizontal strips to reduce the vertical heat conduction through the foil (Supplementary Note 5). To estimate the effective non-radiative heat coefficient of the SHTL region in the vertical direction, we solved the heat transfer equation based on experimentally measured temperature variation along the height (Supplementary Note 6). The extracted $h_{c,SHTL}$ was 1.01 Wm$^{-2}$K$^{-1}$, which was small enough to attain sub-ambient cooling of the TE when $W_{sun}$ < 812 Wm$^{-2}$, as described in Fig. 2c. This $h_{c,SHTL}$ can be easily adjusted by controlling the height of the SHTL, and the filling ratio of the frame on the horizontal surface. Figure 3f shows that $h_{c,SHTL}$ can be lowered by increasing the height and decreasing the filling ratio. A smaller filling ratio also enhances the radiative cooling power per unit area of the TE because the EPS walls block the emission from the TE regions directly underneath them. On the other hand, the mechanical strength of the frame may become a problem if the filling ratio is too small (Supplementary Note 5). Therefore, the structural parameters should be appropriately chosen in practical applications according to the thermal and mechanical requirements of the application.

Finally, we adopted a very simple design for the TE of 330 μm thick polydimethylsiloxane (PDMS) coated on an Al foil. Its emittance in the mid-IR range is close to unity over a broad range of wavelengths. Its absorptance is high in the solar spectral range with $W_{sun}$ = 597 Wm$^{-2}$ under AM1.5 conditions, but this does not result in actual absorption because the radiative interaction of the TE with the outside is spectrally filtered by the SSF. The calculated $W_{TE}$ reached 89 Wm$^{-2}$ at $T_{amb}$ for the red SSF and 87 Wm$^{-2}$ at $T_{amb}$ for the black SSF (Supplementary Note 7).

We included two other SSFs as reference systems to demonstrate the strong dependence of the cooling performance on the spectral selectivity of the SSF. The first reference system was a blackbody (BB) filter that does not transmit but fully absorbs any radiation from UV to mid-IR and is made of black foil (Metal Velvet$^{TM}$, Acktar Ltd.) attached to either side of a Si wafer. The measured absorptance was 99.5 % on average in the wavelength range 0.3–24 μm (Supplementary Note 8). The second reference system was an SSF-less case that is all-transmissive in the solar and mid-IR spectral ranges. The cooling performance of the TE without spectral filtering on top can be checked using this reference. In this case, convection between the outer ambient air and the air of the SHTL may interrupt the cooling performance of the TE, but the thermal state of the TE (whether its temperature is above or below $T_{amb}$) cannot be changed by the ambient air.

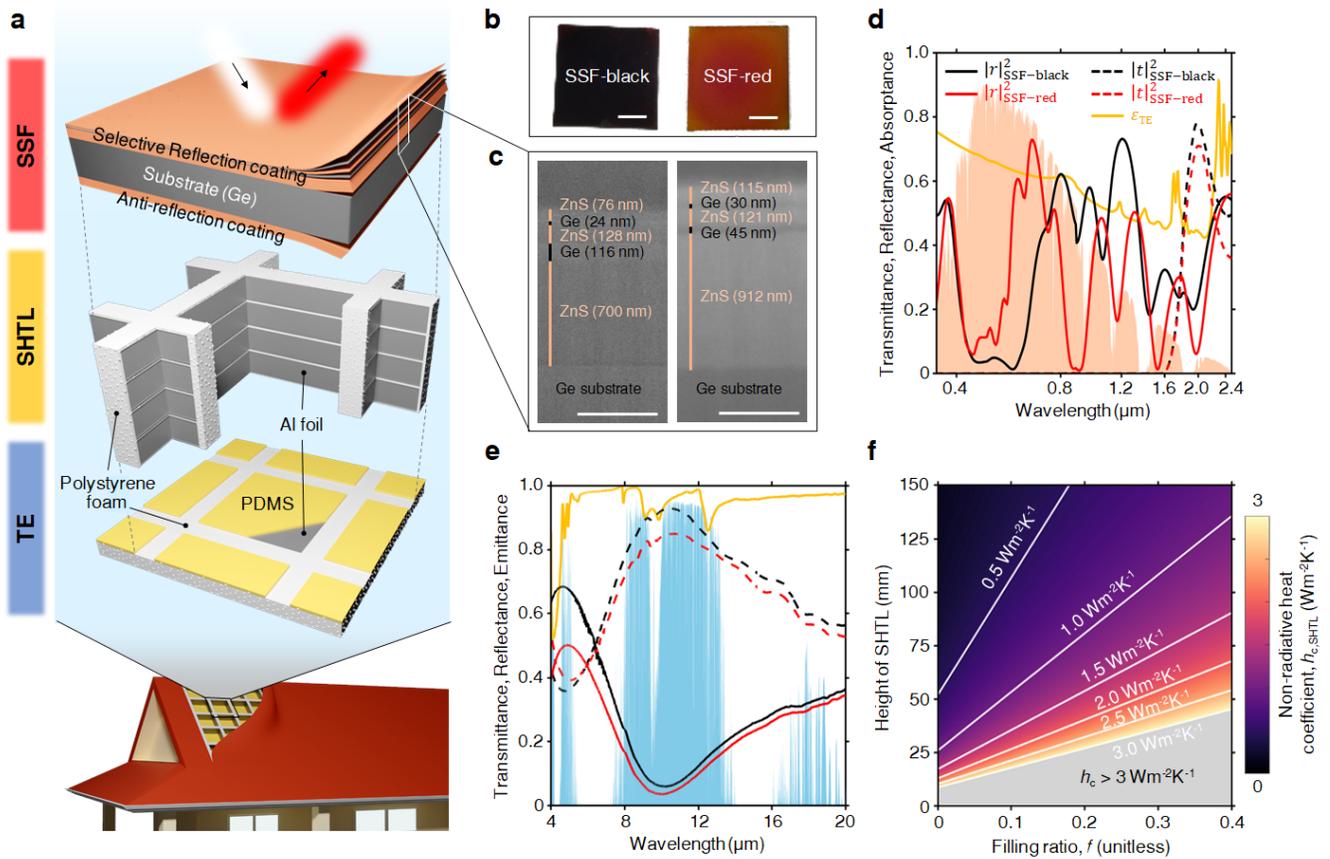

**Fig. 2 Experimental design of RINE system. a** The structural configuration of RINE system, comprising three parts (SSF, SHTL, and TE). The rooftop schematic in the bottom is a conceptual image of how the system is applied in practical situations. **b, c** SSFs for black and red colors: Photographic images (white bar is 10 mm scale) (**b**) and cross-section scanning electron microscopy images (white bar is 500 nm scale) (**c**). **d, e** The reflectance and transmittance profiles of designed SSFs and absorptance (or emittance) of the TE in solar spectral region (**d**) and mid-IR region (**e**). The colored areas in (**d**) and (**e**) represent solar

irradiance and atmospheric transmittance, respectively. **f** Non-radiative heat coefficient of SHTL, $h_{c,\text{SHTL}}$, depending on height and filling ratio of the frame. In this plot, high $h_{c,\text{SHTL}}$ over 3 Wm$^{-2}$K$^{-1}$ is colored in gray for easy illustration.

The outdoor cooling performance measurement setup is shown in Fig. 4a. The samples were placed in separate chambers where the side and bottom were thermally insulated by vacuum-insulated panels, which ensured that the samples' main thermal interactions were approximately one-dimensional as modelled in Eqs. (1) and (2): The outer surface of the setup was covered with barium sulfate paint (pre-mix white coating, Edmund Optics Inc.) to exclude parasitic solar heating of the samples through the setup. In addition, we measured the ambient temperature near the setup where white-coated extruded polystyrene (EPS) shields direct solar illumination. Such a solar shield on the ambient sensor can prevent the overestimation of the ambient temperature and the cooling performance of the samples under test. During the measurements, the setup and pyranometer were placed horizontally on the ground. The outdoor experiment was conducted at KAIST in Daejeon (36.4 °N, 127.4 °E), South Korea, and the climate conditions on the measurement day were as follows: during the daytime, the atmosphere was cloudless, the average wind speed was 3.8 ms$^{-1}$, the average humidity was 57 %, and the solar irradiance reached a maximum value of 743 Wm$^{-2}$ on a horizontal surface (more details are provided in Supplementary Note 9).

**Experimental results.** The outdoor temperature measurement results for the RINE systems are presented in Fig. 4b. In daytime, black and red SSFs were heated up to 14.0 K and 12.5 K over $T_{\text{amb}}$, with 5.3 K and 5.0 K over $T_{\text{amb}}$ on average, respectively. However, the TEs for both the black and red SSFs were cooled below $T_{\text{amb}}$ at all time during the day. Figure 4c shows their temperatures dropped to 6.9 K and 7.4 K below $T_{\text{amb}}$, with 3.5 K and 3.7 K below $T_{\text{amb}}$ on average, respectively. This showed that the RINE scheme can achieve sub-ambient cooling even with dark-colored exteriors directly exposed to the sun on a clear day. This also implied that the color in the SSF can be diversified significantly over what was previously possible with an isothermally colored emitter while keeping the TE cool below $T_{amb}$. To check the possible color variety in the actual experimental conditions, we compared the $W_{sun}$ of the fabricated SSFs with that of all other colors in the absorptive color space as described in Supplementary Note 1. The $W_{sun}$ of SSFs can be calculated using the spectral solar irradiance in the horizontal plane and the absorptive spectrum of SSFs at the solar elevation angle (Supplementary note 10). The calculated $W_{sun}$ values of the black and red SSFs reached peak values of 475 $Wm^{-2}$ and 490 $Wm^{-2}$ at noon, respectively. Interestingly, such values of $W_{sun}$ can cover over 99% of the volume fraction of the three-dimensional absorptive color space at noon in mid-latitude summer (with optimal colorants with minimal $W_{sun}$ for each given color assumed), which is a major advance compared to the previous color expression limit of 5% volume fraction.

On the other hand, the reference system with the BB filter presented quite different results in the daytime. Contrary to the optimized SSFs, the BB filter not only blocked the radiative interaction of the TE with the outer space but also emitted significant thermal radiation to the TE. Consequently, the results show that the TE was heated up to 6.8 K above $T_{amb}$ and stayed above the ambient temperature during high solar altitude time from 10:17 to 15:32. In this case, the non-radiative heat flow was smaller than that of the RINE system because of the smaller temperature gap between the filter and the TE, thus revealing that the overheating of the TE mainly arises from radiative heat from the hot BB filter. Therefore, a good transmittance in the mid-IR atmospheric transparency window is a critical requirement for the SSF if sub-ambient cooling is intended. Without it, the emitter temperature rises above the ambient even when all solar energy is blocked by the SSF.

As another point of reference, we include the results for the all-transmissive SSF case. In that configuration, the solar illumination reached the TE, and the TE was heated up to 4.7 K above $T_{amb}$. It remained above-ambient temperature during most times of the day from 10:02 to 17:27. This emphasizes the importance of the ability of the SSF to block sunlight. The TEs in both reference cases cooled below $T_{amb}$ only at night or near sunrise and sunset when solar illumination is weak due to the low elevation angle of the sun.

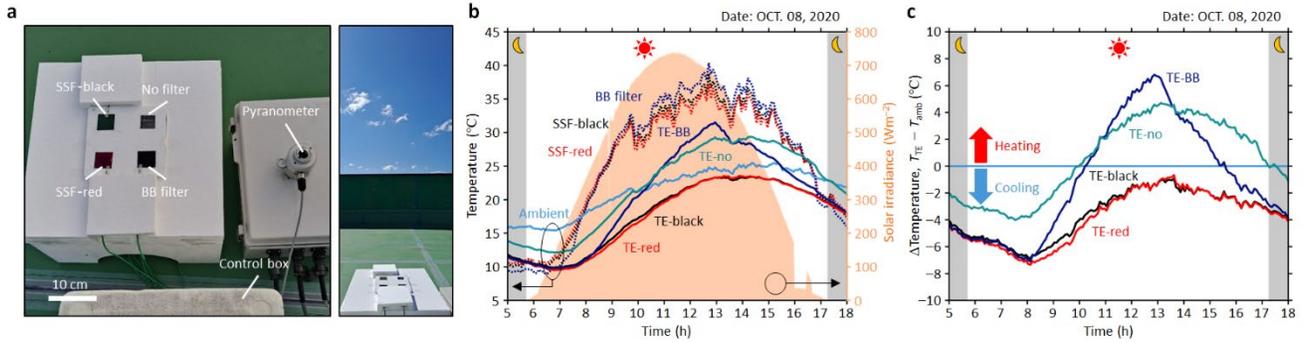

**Fig. 3 Outdoor temperature measurement results. a** Photographic images for measurement setup. The right picture shows the sky conditions on the day the measurement was conducted. **b** The measured temperature for SSFs, TEs, and the ambient. The colored area indicates solar irradiance (scales on the right). **c** The temperature of the TEs compared to $T_{amb}$. The gray-colored area indicates nighttime.

**Conclusions**

In conclusion, we have proposed a non-equilibrium design for colored radiative cooling, which ensures that an internal emitter is able to release radiative heat to outer space but to be insulated from radiative and non-radiative heat sources outdoors. Based on the theoretical study in colorimetry, we have shown that the potential color variety in our scheme is significantly extended beyond the color expression limit in previous designs of colored emitters for sub-ambient cooling, and it is possible to use almost any color in the absorptive color space. In addition, with the help of the spectral filtering ability of the outer layer, a simple broadband emitter can be used instead of a precisely designed wavelength-selective emitter. We have experimentally demonstrated RINE systems for black and dark-red colors with broadband emitters that are cooled below the ambient temperature during the daytime. Our design suggests a viable way to express diverse colors without severe degradation of radiative cooling performance, potentially increasing the commercial feasibility of radiative cooling for wider range of practical applications.

## Methods

**Calculation of the net cooling power density.** To solve the coupled Eqs. (1) and (2), we enumerate the radiative and non-radiative heat exchange terms between subsystems and between a subsystem and the environment. First, the net radiative cooling power densities of the SSF and the TE can be arranged as

$$W_{\text{SSF}} = W_{\text{SSF,all}} - W_{\text{TE,SSF}} - W_{\text{sun,SSF}} - W_{\text{atm,SSF}} - W_{\text{space,SSF}}, \quad (3)$$

$$W_{\text{TE}} = W_{\text{TE,all}} - W_{\text{SSF,TE}} - W_{\text{sun,TE}} - W_{\text{atm,TE}} - W_{\text{space,TE}}, \quad (4)$$

where $W_{\text{SSF,all}}$, and $W_{\text{TE,all}}$ are the radiant exitance of the SSF and TE, respectively, excluding the amount absorbed back by themselves due to reflections from other surfaces; $W_{\text{TE,SSF}}$, and $W_{\text{SSF,TE}}$ are the absorbed irradiances on the SSF, and TE emitted from the TE, and SSF, respectively; $W_{\text{sun,SSF}}$, and $W_{\text{sun,TE}}$ are the absorbed irradiances on the SSF and TE, respectively, emitted from the sun; $W_{\text{atm,SSF}}$, and $W_{\text{atm,TE}}$ are the absorbed irradiances on the SSF and TE, respectively, emitted from the atmosphere; and $W_{\text{space,SSF}}$, and $W_{\text{space,TE}}$ are the absorbed irradiances on the SSF and TE, respectively, emitted from outer space. In this case, $W_{\text{space,SSF}}$ and $W_{\text{space,TE}}$ can be ignored owing to the large radiative temperature difference of outer space (~3 K) and other objects. Given the radiative properties of the subsystems and environment, the remaining terms in Eqs. (3) and (4) can be expressed as

$$W_{\text{SSF,all}}(T_{\text{SSF}}) = \iint \left[ \tilde{I}_{\text{BB}}(\lambda, T_{\text{SSF}}) \left( \varepsilon_{\text{SSF,1}}(\lambda, \Omega, T_{\text{SSF}}) + \varepsilon_{\text{SSF,2}}(\lambda, \Omega, T_{\text{SSF}}) - \frac{|r_{\text{TE}}(\lambda,\Omega,T_{\text{TE}})|^2 \varepsilon_{\text{SSF,2}}^2(\lambda,\Omega,T_{\text{SSF}})}{1-|r_{\text{TE}}(\lambda,\Omega,T_{\text{TE}})r_{\text{SSF,2}}(\lambda,\Omega,T_{\text{SSF}})|^2} \right) \right] \cos\theta \, d\lambda d\Omega,$$
(3a)

$$W_{\text{TE,SSF}}(T_{\text{SSF}}, T_{\text{TE}}) = \iint \left[ \tilde{I}_{\text{BB}}(\lambda, T_{\text{TE}}) \frac{\varepsilon_{\text{TE}}(\lambda,\Omega,T_{\text{TE}}) \varepsilon_{\text{SSF,2}}(\lambda,\Omega,T_{\text{SSF}})}{1-|r_{\text{TE}}(\lambda,\Omega,T_{\text{TE}})r_{\text{SSF,2}}(\lambda,\Omega,T_{\text{SSF}})|^2} \right] \cos\theta \, d\lambda d\Omega, \quad (3b)$$

$$W_{\text{sun,SSF}}(T_{\text{SSF}}, T_{\text{TE}}) = \iint \left[ \tilde{I}_{\text{sun}}(\lambda, \Omega) \left( \varepsilon_{\text{SSF,1}}(\lambda, \Omega, T_{\text{SSF}}) + \frac{|r_{\text{TE}}(\lambda,\Omega,T_{\text{TE}}) t_{\text{SSF}}(\lambda,\Omega,T_{\text{SSF}})|^2 \varepsilon_{\text{SSF,2}}(\lambda,\Omega,T_{\text{SSF}})}{1-|r_{\text{TE}}(\lambda,\Omega,T_{\text{TE}})r_{\text{SSF,2}}(\lambda,\Omega,T_{\text{SSF}})|^2} \right) \right] \cos\theta \, d\lambda d\Omega, \quad (3c)$$

$$W_{\text{atm,SSF}}(T_{\text{SSF}}, T_{\text{TE}}, T_{\text{amb}}) = \iint \left[ \tilde{I}_{\text{atm}}(\lambda, \Omega, T_{\text{amb}}) \left( \varepsilon_{\text{SSF,1}}(\lambda, \Omega, T_{\text{SSF}}) + \frac{|r_{\text{TE}}(\lambda,\Omega,T_{\text{TE}}) t_{\text{SSF}}(\lambda,\Omega,T_{\text{SSF}})|^2 \varepsilon_{\text{SSF,2}}(\lambda,\Omega,T_{\text{SSF}})}{1-|r_{\text{TE}}(\lambda,\Omega,T_{\text{TE}})r_{\text{SSF,2}}(\lambda,\Omega,T_{\text{SSF}})|^2} \right) \right] \cos\theta \, d\lambda d\Omega,$$
(3d)

$$W_{\text{TE,all}}(T_{\text{TE}}) = \iint \left[ \tilde{I}_{\text{BB}}(\lambda, T_{\text{TE}}) \left( \varepsilon_{\text{TE}}(\lambda, \Omega, T_{\text{TE}}) - \frac{|r_{\text{SSF,2}}(\lambda,\Omega,T_{\text{SSF}})|^2 \varepsilon_{\text{TE}}^2(\lambda,\Omega,T_{\text{TE}})}{1-|r_{\text{TE}}(\lambda,\Omega,T_{\text{TE}})r_{\text{SSF,2}}(\lambda,\Omega,T_{\text{SSF}})|^2} \right) \right] \cos\theta \, d\lambda d\Omega, \quad (4a)$$

$$W_{\text{SF,TE}}(T_{\text{SSF}}, T_{\text{TE}}) = \iint \left[ \tilde{I}_{\text{BB}}(\lambda, T_{\text{SSF}}) \frac{\varepsilon_{\text{TE}}(\lambda,\Omega,T_{\text{TE}}) \varepsilon_{\text{SSF,2}}(\lambda,\Omega,T_{\text{SSF}})}{1-|r_{\text{TE}}(\lambda,\Omega,T_{\text{TE}})r_{\text{SSF,2}}(\lambda,\Omega,T_{\text{SSF}})|^2} \right] \cos\theta \, d\lambda d\Omega, \quad (4b)$$

$$W_{\text{sun,TE}}(T_{\text{SSF}}, T_{\text{TE}}) = \iint \left[ \tilde{I}_{\text{sun}}(\lambda, \Omega) \frac{|t_{\text{SSF}}(\lambda,\Omega,T_{\text{SSF}})|^2 \varepsilon_{\text{TE}}(\lambda,\Omega,T_{\text{TE}})}{1-|r_{\text{TE}}(\lambda,\Omega,T_{\text{TE}})r_{\text{SSF,2}}(\lambda,\Omega,T_{\text{SSF}})|^2} \right] \cos\theta \, d\lambda d\Omega, \quad (4c)$$

$$W_{\text{atm,TE}}(T_{\text{SSF}}, T_{\text{TE}}, T_{\text{amb}}) = \iint \left[ \tilde{I}_{\text{atm}}(\lambda, \Omega, T_{\text{amb}}) \frac{|t_{\text{SSF}}(\lambda, \Omega, T_{\text{SSF}})|^2 \varepsilon_{\text{TE}}(\lambda, \Omega, T_{\text{TE}})}{1 - |r_{\text{TE}}(\lambda, \Omega, T_{\text{TE}}) r_{\text{SSF},2}(\lambda, \Omega, T_{\text{SSF}})|^2} \right] \cos\theta \, d\lambda d\Omega, \quad (4d)$$

where $\int (\cdot) d\Omega = \int_0^{2\pi} \int_0^{\pi/2} (\cdot) \sin\theta \, d\theta \, d\phi$ is the hemispherical integration with solid angle $\Omega$ and the functions and parameters in the integrands are as follows: $\tilde{I}_{\text{BB}}$, $\tilde{I}_{\text{sun}}$, and $\tilde{I}_{\text{atm}}$ (Wm$^{-2}$μm$^{-1}$sr$^{-1}$) are the spectral radiances of a blackbody, the sun, and the atmosphere, respectively; $T_{\text{SSF}}$, $T_{\text{TE}}$, and $T_{\text{amb}}$ are the temperatures of the SSF, TE, and ambient air, respectively; $|r_{\text{SSF},1}|^2$ and $|r_{\text{SSF},2}|^2$ are the reflectances for the outer and inner sides of the SSF, respectively; $|r_{\text{TE}}|^2$ is the reflectance of the TE; $\varepsilon_{\text{SSF},1}$ and $\varepsilon_{\text{SSF},2}$ are the emittances of the outer and inner sides of the SSF, respectively; $\varepsilon_{\text{TE}}$ is the emittance of the TE; $|t_{\text{SSF}}|^2$ is the transmittance of the SSF. Because of reciprocity, the transmittances in both directions through the SSF are the same. The spectral radiance of a blackbody at $T$ is $\tilde{I}_{\text{BB}}(\lambda, T) = 2hc^2 \lambda^{-5}[\exp(hc\lambda^{-1}k_B^{-1}T^{-1}) - 1]^{-1}$ where $h$ is Planck's constant, $c$ is the speed of light in a vacuum, and $k_B$ is the Boltzmann constant. We obtain $\tilde{I}_{\text{sun}}(\lambda, \Omega)$ for given temporal and spatial conditions by using a simple model of the atmospheric radiative transfer of sunshine (SMARTS 2.9.5, NREL). We calculated $\tilde{I}_{\text{atm}}(\lambda, \Omega, T_{\text{amb}})$ by multiplying $\tilde{I}_{\text{BB}}(\lambda, T_{\text{amb}})$ and the atmospheric emittance $\varepsilon_{\text{atm}}(\lambda, \Omega, T_{\text{amb}})$ which was modeled by a spherical shell model as $\varepsilon_{\text{atm}}(\lambda, \Omega, T_{\text{amb}}) = 1 - t_0(\lambda, T_{\text{amb}})^{\text{AM}(\theta)}$ where $t_0$ is the atmospheric transmittance in the zenith direction, and AM($\theta$) is an attenuation factor of the atmosphere at zenith angle $\theta$[31,32].

In Eq. (3a), the first and second terms are the thermal emissions of the SSF through the outer and inner sides, respectively. The third term accounts for the partial absorption by the SSF itself of the second term owing to multiple reflections between the SSF and the TE. The second term in Eqs. (3c) and (3d) also correspond to additional absorption due to multiple reflections in the cooler. Equation (4a) presents the emitted thermal radiation from the TE. Terms in Eqs. (4a) to (4d) are similarly defined and calculated for the TE instead of the SSF.

Non-radiative power densities in Eqs. (1, 2) are also defined to be positive in the direction from the first to the second object tagged with the subscripts and assumed to have the following linear dependence on the temperature differences:

$$\phi_{\text{SSF,atm}} = h_{c,\text{amb}}(T_{\text{SSF}} - T_{\text{amb}}), \tag{5}$$

$$\phi_{\text{SSF,SHTL}} = -\phi_{\text{TE,SHTL}} = h_{c,\text{SHTL}}(T_{\text{SSF}} - T_{\text{TE}}), \tag{6}$$

where $h_{c,\text{amb}}$ and $h_{c,\text{SHTL}}$ (Wm$^{-2}$K$^{-1}$) are the effective non-radiative heat coefficients of the interface between the SSF and the environment, and the SHTL in the vertical direction. As we assume large-area applications with no apparent temperature gradient in the lateral directions, non-radiative heat flow in the horizontal direction is ignored. $\phi_{\text{SSF,SHTL}}$ and $\phi_{\text{TE,SHTL}}$ have the same magnitude and opposite signs because we assumed that the SHTL does not absorb or emit thermal radiation and is in a quasi-steady state with heat flux that is almost divergence-free. We note that $h_{c,\text{SHTL}}$ is finite in non-isothermal systems; otherwise, the system reverts to isothermal designs (Supplementary Note 3).

**Sample optimization.** We designed SSF-red and SSF-black using a needle optimization method[33], which optimizes the thickness and material choice for each layer of the five-layer stack on a germanium substrate. We chose germanium and zinc sulfide as layer materials because of their transparency and large differences in their refractive indices. The needle optimization method iteratively searches for the optimal layer configuration by calculating the reflectance spectrum $|r(\lambda)|^2$ in each iteration until the cost function $\Delta E(|r(\lambda)|^2)$ converges to a small value. We define $\Delta E(|r(\lambda)|^2)$ to quantify the color mismatch from the target color (black and red) as well as the deviation of the transmittance spectrum from the ideal profile in Fig. 1(d):

$$\Delta E(|r(\lambda)|^2) = \alpha_1 \Delta E^*_{00}(|r(\lambda)|^2, L^*, a^*, b^*) + \alpha_2 \int_{8\,\mu m}^{13\,\mu m}(1 - |r(\lambda)|^2)d\lambda + \alpha_3 \left[\int_{4\,\mu m}^{8\,\mu m}|r(\lambda)|^2 d\lambda + \int_{13\,\mu m}^{20\,\mu m}|r(\lambda)|^2 d\lambda\right] \tag{7}$$

where $\Delta E^*_{00}$ is a color distance metric defined as in CIE DE2000[34] under D65 illumination. We apply normalization factors $\alpha_1$, $\alpha_2$, and $\alpha_3$ to balance the different objectives. While the performance can further increase with more layers, we chose to deposit five layers because they show good performance and are easy to fabricate.

**Data availability**

The original data that support the finding of this research are available from the corresponding author upon reasonable request.

**Acknowledgements**

This work was supported by the National Research Foundation of Korea (NRF) Grant funded by the Korea government (MSIT) (NO. 2018M3D1A1058998 and No. 2021R1A2C2008687).

**Author contributions**

S.J and J.S conceived the study and wrote the manuscript with support from all authors. S.J and S.M performed calculations and modeling of target samples. S.J and S.S conducted fabrications and measurements with support from H.P and H.L. The project was supervised by J.S.

**Competing interests**

The authors declare no competing interests.

Supplementary material

# Daylong sub-ambient radiative cooling with full color exterior

Suwan Jeon[1], Soomin Son[2], Seokhwan Min[1], Hyunjin Park[1], Heon Lee[2], and Jonghwa Shin[1*]

[1]*Department of Materials Science and Engineering, Korea Advanced Institute of Science and Technology, Daejeon 34141, Republic of Korea*

[2]*Department of Materials Science and Engineering, Korea University, Anam-ro 145, Seongbuk-gu, Seoul, 02841, Republic of Korea*

*\* qubit@kaist.ac.kr*

**Note S1: The lower bound of solar thermal load of target color**

**Note S2: Various spectral designs for RINE system**

**Note S3: Coupled equations for extreme $h_{c,SHTL}$**

**Note S4: RINE system depending on $h_{c,amb}$**

**Note S5: Thermal and physical properties according to SHTL structure**

**Note S6: $h_{c,SHTL}$ extraction based on experimental data**

**Note S7: Cooling performance of designed RINE system**

**Note S8: Spectral property of BB filter**

**Note S9: Climate conditions in measurement day**

**Note S10: Spectral angular properties of SSF and TE**

**Note S1: The lower bound of solar thermal load of target color**

We show how the boundary of solar thermal load ($W_\text{sun}$), particularly for the lower side, of target color can be obtained by metamer optimization method[1,2]. In general, the perceived color is quantitatively defined in CIE XYZ color space where a color coordinate $\vec{X}_c$, also referred to as tristimulus values, is determined by color matching functions ($\bar{\mathbf{x}}$: $3 \times k$ matrix in which the rows correspond to the color matching functions ($\bar{x}$, $\bar{y}$, and $\bar{z}$) with $k$ wavelength points), background illuminant ($\mathbf{D}$: $k \times k$ diagonal matrix in which diagonal element accords with the intensity of the illuminant at each wavelength point), spectral reflectance ($\vec{R}_c$: $k \times 1$ matrix):

$$\vec{X}_c = \frac{100}{N}\bar{\mathbf{x}}\mathbf{D}\vec{R}_c \qquad (S1\text{-}1)$$

where $N$ is a spectral integration of multiplication of the illuminant and $\bar{y}$. Usually, the number of sensible wavelengths ($k$) is enormously larger than three of tristimulus values, indicating that $\vec{X}_c$ is highly underdetermined. Thus, one can find color-invariant vectors $\vec{R}_0$ for a given illuminant, satisfying that

$$0 = \frac{100}{N}\bar{\mathbf{x}}\mathbf{D}\vec{R}_0, \qquad (S1\text{-}2)$$

which is called metamer black, does not affect color perception by representing pure black. Combining Eqs. (S1-1) and (S1-2), one can find a general reflectance spectrum $\vec{R}$ of target color $\vec{X}_c$[1] as

$$\vec{R} = \mathbf{A}^T(\mathbf{A}\mathbf{A}^T)^{-1}\vec{X}_c + \mathbf{B}\vec{\alpha} \qquad (S1\text{-}3)$$

where $\mathbf{A} = \bar{\mathbf{x}}\mathbf{D}$, $\mathbf{B}$ is a $k \times (k-3)$ matrix in which the columns correspond to the metamer black basis, and $\vec{\alpha}$ is a weighting vector of $k-3$ length. In Eq. (S1-3), different $\vec{\alpha}$ may change $\vec{R}$ but the tristimulus value ($\vec{X}_c$) remains the same. When the transmission of the colored object is zero, the absorbed irradiance can be calculated from the rest of the reflected illuminant as

$$W_\text{sun} = \vec{u}\mathbf{D}(1 - \vec{R}) \qquad (S1\text{-}4)$$

where $\vec{u}$ is a 1 × $k$ vector in which all element is a unit, which is implemented for spectral summation.

Then, one can find the lower (or upper) bound of $W_{sun}$ by optimizing $\vec{a}$ of color specified $\vec{R}$. Meanwhile, the half of solar power is ranged outside the visible region, such as UV and near-IR, indicating that the absorption in the unperceivable region should be suppressed (allowed) for the minimal (maximal) absorption.

Based on this approach, we illustrate a possible range of $W_{sun}$ in conventional radiative surfaces for various colors (Fig. S1). In this case, the illuminant is a solar irradiance at noontime of mid-latitude summer. The results in Fig. S1a show that $W_{sun}$ for most colors (such as black, red, blue, green magenta, and yellow) fundamentally exceeds $W_{max,300}$, indicating that such colors never reach sub-ambient temperature via the radiative cooling process. Interestingly, it also reveals that a perfect white can be radiatively heated over $T_{amb}$ when solar absorption in the invisible region is high (Fig. S1b). In Fig. S1c, one can check the lower bound of $W_{sun}$ for all colors in the sRGB gamut, presenting that only 5% of colors wherein high lightness ($L$) and low saturation ($\sqrt{a^2 + b^2}$) are available for all-time sub-ambient cooling.

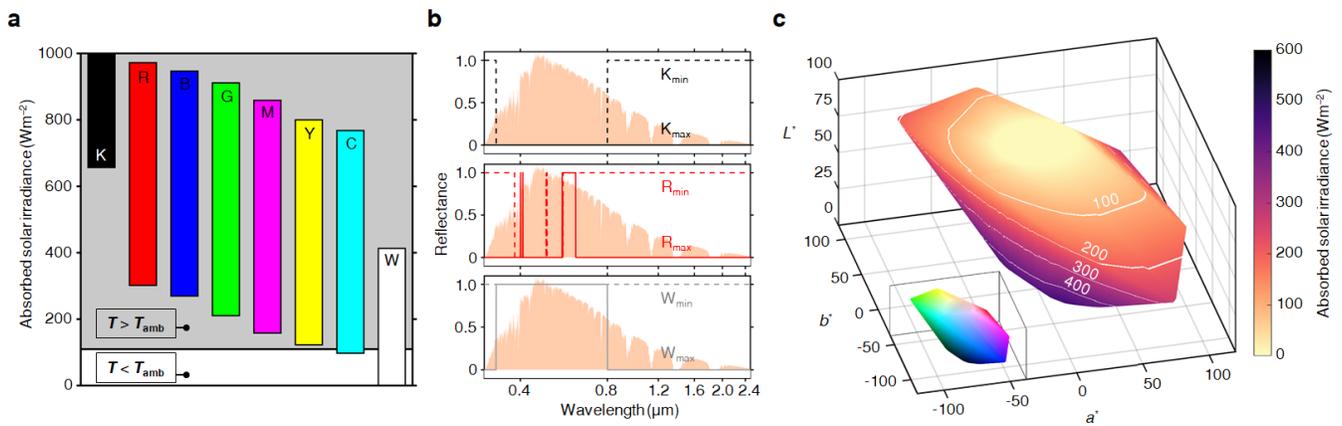

**Fig. S1 Solar absorption to generate colors**. (**a**) The possible range of absorbed solar irradiance to generate representative colors (K: black, R: red, B: blue, G: green, M: magenta, Y: yellow, C: cyan, W: white) when the surface is horizontally placed on the ground at noontime of mid-latitude summer. The gray and white areas present the above-ambient and sub-ambient temperature regions. (**b**) The spectral reflectance of specific color whose solar thermal load corresponds to the upper bound (solid line, $X_{max}$) or the lower bound (dashed line, $X_{min}$). The colored area indicates the solar irradiance with a scaled value. (**c**) The lower bound of absorbed solar irradiance for all perceivable colors in the sRGB gamut. The inset illustrates a color map on a smaller scale.

**Note S2: Various spectral designs for RINE system**

We examine the radiative cooling performance for various spectral designs of SSF and TE (Fig. S2). First, we introduce SSF of type A, type B, and type C whose photonic properties are reflective, transmissive, and absorptive in the mid-infrared range except in 8–13 μm wavelengths that accord with the radiative cooling passage. We note that all types of SSFs do not allow transmission at below 4 μm. For TE, we implement well-known emitters whose emittances are broadband over 4 μm, selective in 8–13 μm, and ideal at specific wavelengths, where SSF allows the transmission, for the maximum of radiative cooling power[3]. In the meantime, SHTL in the middle allows radiations to pass in either direction. Considering the different spectral designs above, one can obtain the steady-state temperatures of SSF and TE by solving the coupled power equations in Methods. The results in Fig. S2b show that, when SSF and TE are radiatively coupled (e.g., broadband TE and SSF of type A), intense radiative heat from hot SSF is transferred to TE and substantially degrades the cooling performance of TE. On the other hand, when SSF and TE are radiatively decoupled, TE can reach sub-ambient temperature with small $h_{c,SHTL}$. This also notes that non-radiative decoupling between TE and SSF is essential for efficient radiative cooling. To check how ample non-radiative insulation is necessary for all-time sub-ambient cooling with arbitrary color expression, we present $h_{c,SHTL}$ corresponding to TE at $T_{amb}$ while absorbing 1000 Wm$^{-2}$ of sunlight in Table S1. The results show that, among various RINE systems, the combination of ideal (or selective) TE and SSF of type C present the largest $h_{c,SHTL}$ (i.e., least non-radiative insulation) to obtain sub-ambient cooling, which indicates the thinnest SHTL for the same insulation.

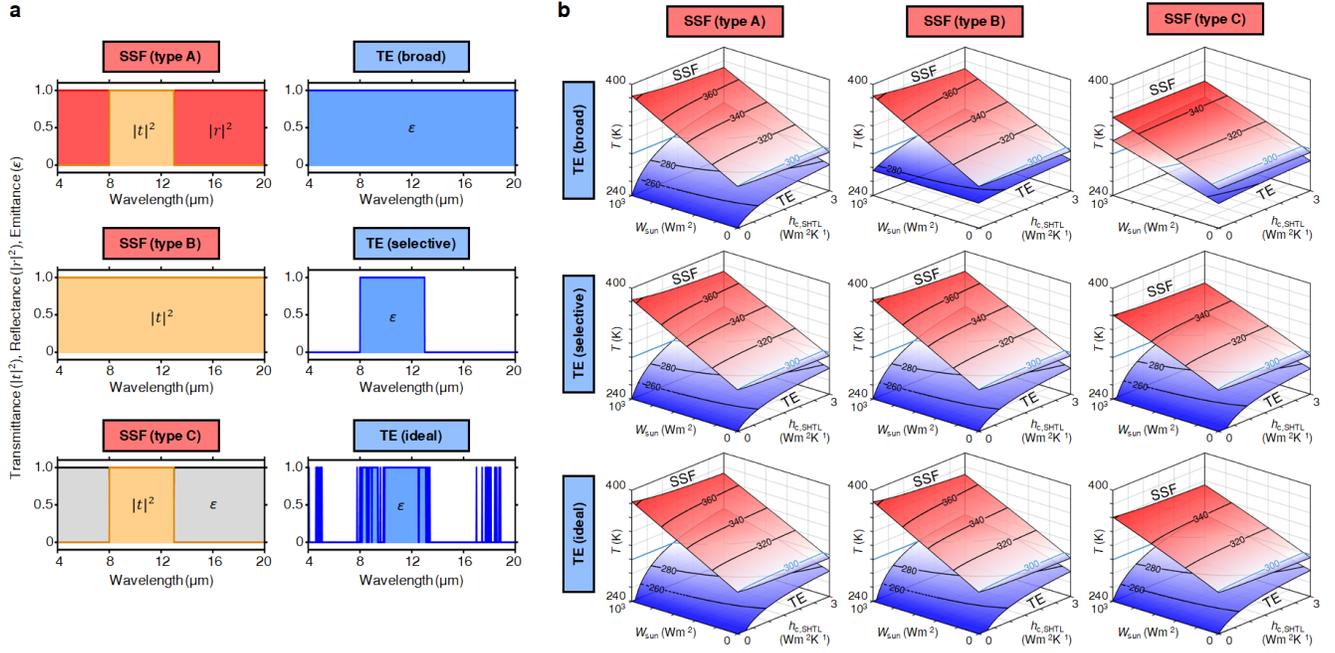

**Fig. S2 The steady-state temperature of the RINE system for various spectral combinations.** (**a**) Spectral design of SSFs (type A, type B, and type C) and TEs (broad, selective, and ideal types). The ideal spectrum of TE is designed only at the wavelengths where matched with the transmission region of SSF. (**b**) The steady-state temperatures of SSF and TE depending on $h_{c,SHTL}$, and $W_{sun}$, for various combinations. In this case, the absorptance (emittance) of TE is zero at wavelengths below 4 μm to prevent solar heating of TE. Also, $h_{c,amb}$ is constant as 12 Wm$^{-2}$K$^{-1}$.

| | | | Unit: Wm$^{-2}$K$^{-1}$ |
|---|---|---|---|
| | SSF (type A) | SSF (type B) | SSF (type C) |
| TE (broad) | 1.35 | 1.48 | - |
| TE (selective) | 1.35 | 1.35 | 1.83 |
| TE (ideal) | 1.35 | 1.49 | 1.83 |

**Table S1. Non-radiative heat coefficient of SHTL for various spectral combinations of SSF and TE, when TE is at $T_{amb}$ and SSF absorbs 1000 Wm$^{-2}$ of sunlight.**

**Note S3: Coupled equations for extreme $h_{c,\text{SHTL}}$**

We show that, for extreme $h_{c,\text{SHTL}}$, the coupled equations of Eqs. (1, 2) in the main text can be rearranged into the equation for a well-known isothermal system. If $h_{c,\text{SHTL}}$ is large enough to make $\phi_{\text{SHTL,SSF}}$ and $\phi_{\text{SHTL,TE}}$ dominate the other terms in Eqs. (1, 2), the temperature differences of SSF and TE will be negligible at the steady-state. Then, one can integrate the equations (1) and (2) with an isothermal temperature $T_{\text{uni}}$ (= $T_{\text{SSF}} = T_{\text{TE}}$) by canceling out $\phi_{\text{SSF,SHTL}}$ and $\phi_{\text{TE,SHTL}}$ as

$$P_{\text{iso}} = P_{\text{SSF}} + P_{\text{TE}} = W_{\text{SSF}} + W_{\text{TE}} + \phi_{\text{SSF,atm}} \quad \text{(S3-1)}$$

where $P_{\text{iso}}$ is a net cooling power density of the unified system. As disclosed in the Eqs. (3, 4), $W_{\text{SSF}}$ and $W_{\text{TE}}$ are the net radiatively emitted power densities of each subsystem, and their radiative interplays have the opposite sign. Therefore, the summation can annul radiative interplay between SSF and TE, and present the net radiative cooling power density of the unified system. Also, $\phi_{\text{SSF,atm}}$ represents the non-radiative power density of the isothermal system. Therefore, the equation (S3-1) expresses the same equation for the isothermal radiative cooling system in previous studies[4].

**Note S4: RINE system depending on $h_{c,amb}$**

Now, we demonstrate how the atmospheric non-radiation affects the thermodynamic state of RINE system. In Fig. S3, we illustrate the steady-state temperature of SSF and TE, whose spectral properties follow the designs in the manuscript, depending on $W_{sun}$ and $h_{c,amb}$. The results exhibit that, for large $h_{c,amb}$, the temperature of SSF is substantially lowered by dissipating non-radiative heat into the relatively cold ambient. Then, the temperature difference between SSF and TE is reduced, indicating less non-radiative heat flow from SSF to TE. Following this process, TE is further cooled for the same non-radiative insulation of SHTL as $h_{c,amb}$ increases. Considering that $h_{c,amb}$ is usually larger than 6 Wm$^{-2}$K$^{-1}$ in outdoors[5], $h_{c,SHTL}$ of 0.5 Wm$^{-2}$K$^{-1}$ can sustain the sub-ambient temperature of TE while allowing perfect solar absorption in SSF (Fig. S3a). Even under six times larger $h_{c,SHTL}$ condition as 3 Wm$^{-2}$K$^{-1}$, sub-ambient cooling of TE is possible with the absorption of half of the sunlight in SSF, which implies that various colors in Fig. S1 are available except for black.

We also note two critical differences compared to previous radiative coolers. First, enhancing $h_{c,amb}$ can change TE from the above-ambient state to the sub-ambient state, which has never been possible in the isothermal system by simply adjusting $h_{c,amb}$. Similar to this, $h_{c,SHTL}$ can also improve radiative cooling performance as shown in the manuscript. Practically, such additional factors to control the thermal state can be helpful to design the sub-ambient cooler. Second, more importantly, increasing $h_{c,amb}$ degrades the cooling performance of the isothermal emitter contrary to RINE system. So, previous works have extensively studied to lower $h_{c,amb}$ by implementing non-radiative shields, such as thin polymeric sheet[4,6] and aerogel[7,8] on the above the cooler. However, due to the opposite function of atmospheric non-radiation for sub-ambient cooling, our RINE system does not require previous insulation kits and can operate as a standalone system.

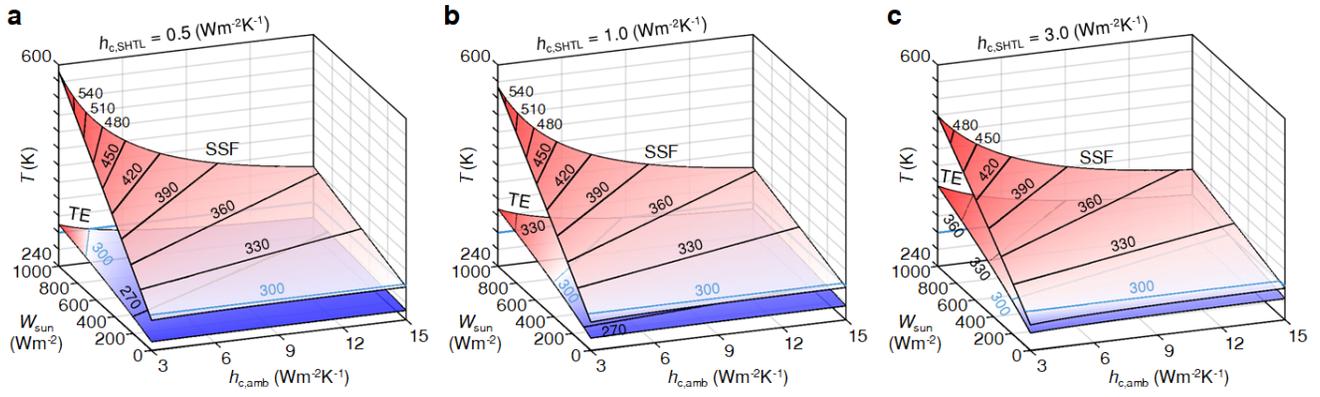

**Fig. S3 The steady-state temperature of SSF and TE depending on $W_{sun}$ and $h_{c,amb}$.** In each case, $h_{c,SHTL}$ is (**a**) 0.5 Wm$^{-2}$K$^{-1}$, (**b**) 1.0 Wm$^{-2}$K$^{-1}$, and (**c**) 3.0 Wm$^{-2}$K$^{-1}$.

**Note S5: Thermal and physical properties according to SHTL structure**

In SHTL, aluminum (Al) mirrors are attached to polystyrene (PS) walls to transfer the radiations from TE to SSF. However, the heat conductivity of Al is much larger than that of the rest materials in SHTL. So, we study how much conductive heat flows through the mirror layer according to the mirror structure. First, we consider SHTL composed of a uniform mirror layer on PS, and the rest is filled with air (Fig. S4a). As an equivalent circuit, it can be represented as parallel resistors of $R_{\text{air}} = HA_{\text{air}}^{-1}k_{\text{c,air}}^{-1}$, $R_{\text{PS}} = HA_{\text{PS}}^{-1}k_{\text{c,PS}}^{-1}$, and $R_{\text{mirror}} = HA_{\text{mirror}}^{-1}k_{\text{c,mirror}}^{-1}$ where $H$ is the height of SHTL, $A_{\text{air}}$, $A_{\text{PS}}$, $A_{\text{mirror}}$ are the area of the air, PS, and the mirror, $k_{\text{c,air}}$, $k_{\text{c,PS}}$, $k_{\text{c,mirror}}$ are the thermal conductivities of the air, PS, and the mirror. When SHTL is periodic, one can use the filling ratio ($f$) of each material instead of the area (or volume). Then, depending on the filling ratio of the mirror, one can calculate the thermal conductance ratio of the mirror for given $f_{\text{PS}}$; in this case, the thermal conductance is the inverse of the thermal resistance. The result in Fig. S4c shows that most of the conductive heat flows through the mirror layer even at small $f_{\text{mirror}}$ less than 1% due to large $k_{\text{c,mirror}}$. To suppress heat flow through the mirror, $f_{\text{mirror}}$ needs to be reduced, but at the same time, the thickness of the mirror layer should be thick enough (~100 nm) to reflect thermal radiation. However, when the period of SHTL is 5 cm and $f_{\text{PS}}$ is 10%, the thickness of the mirror should be less than ~30 nm to achieve 1% of the thermal conductance ratio of the mirror. This result indicates that the uniform metallic mirror is challenging to realize low thermal conductivity and high mid-IR reflectance together.

As an alternative, we can implement divided mirror layers in Fig. S4b. The multiple mirror layers are a series of the air and the mirror parts, but near the air gap, there can be local conductive heat flow through side PS walls (inset of Fig. S4b). Then, the total resistance can be represented as $L_{\text{gap}}(L_{\text{gap}}+L_{\text{mirror}})^{-1}R_{\text{air}}\|R_{\text{PS}} + L_{\text{mirror}}(L_{\text{gap}}+L_{\text{mirror}})^{-1}R_{\text{mirror}}$ where $L_{\text{gap}}$ and $L_{\text{mirror}}$ are the total lengths of the air gap and the mirror. The local conductive heat flow in the gap depends on structural design, so it is difficult to be represented simply by $f_{\text{mirror}}$. Instead, one can set the thermal conductance range to account for the cases in which the gaps in the mirror are filled with air or PS. Figure S4d shows the possible range of thermal conductance ratio of the composite mirror for $L_{gap} = 0.01L_0$ and $L_{mirror} = 0.99L_0$. This result presents that the air gap significantly reduces heat flow through the divided mirror compared to the uniform mirror. Also, for 1% of thermal conductance ratio of the mirror, the thickness of the divided mirror is ranged in 244−2490 nm, which is much thicker than that of the uniform mirror (~30 nm) and allows to reflect almost all radiations; when the period of SHTL is 5 cm and $f_{PS}$ is 10%.

Additionally, we demonstrate the radiative cooling performance and physical strength depending on SHTL. Figure S5a shows two possible structures for SHTL, which can be described by the filling ratio of the composite walls, $f_{wall}$. When $f_{wall}$ is small, the area of TE (i.e. $1 - f_{wall}$) relatively increases, so the radiative cooling power per unit cell is enhanced, as shown in Fig. S5b. Also, as $f_{wall}$ decreases, the filling ratio of the air increases, which strengthens the thermal insulation of SHTL (Fig. 3f). Based on these results, one can calculate the steady-state temperature of TE as illustrated in Fig. S5c; in this case, $T_{SSF} = 320$ K, and the thermal conductance through the mirror surface is ignored. These results note that, for the same height, the cooling performance of TE is improved when $f_{wall}$ is small. In the meantime, the physical strength of SHTL is another essential factor that must be considered for practical implementation. To check this, Young's modulus ($E_{zz}$) of grid and pillar structures depending on $f_{wall}$ is calculated from COMSOL simulation (Fig. S5d). The simulation results indicate that grid SHTL is more resistant to compression than pillar SHTL, and their $E_{zz}$ is commonly lowered as $f_{wall}$ decreases. Then, for $f_{wall} = 10\%$, $E_{zz}$ of grid SHTL is 7 MPa, which indicates that the grid SHTL can support 71.4 gmm$^{-2}$ for the compressive strain of 10%.

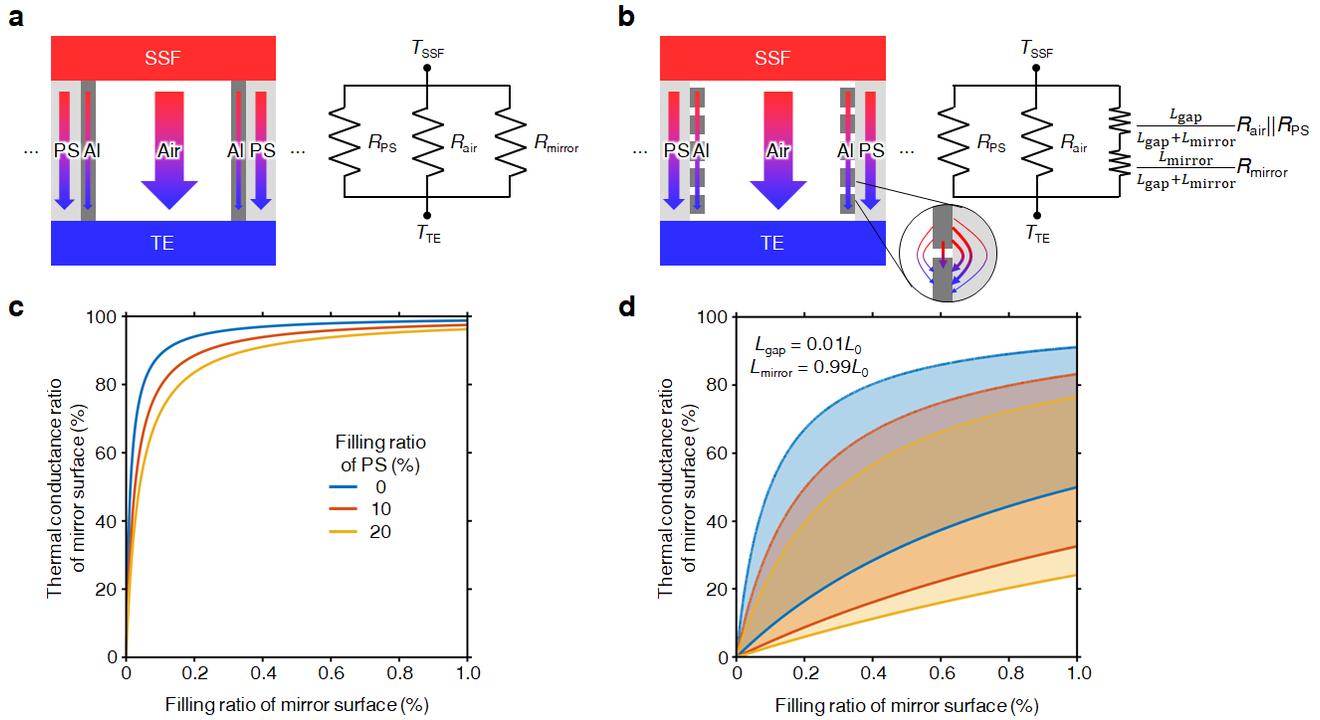

**Fig. S4 Heat conductance of SHTL according to mirror design.** Schematics and circuit models of SHTL composed with (**a**) uniform mirror surface and (**b**) divided mirror surfaces. In circuit models, thermal resistances are divided into three parts: polystyrene ($R_{PS}$), air ($R_{air}$), and mirror parts ($R_{mirror}$ in (a), and $L_{gap}(L_{gap} + L_{mirror})^{-1}R_{air} \| R_{PS} + L_{mirror}(L_{gap} + L_{mirror})^{-1}R_{mirror}$ in (b)). Thermal conductance ratio of (**c**) uniform mirror surface and (**d**) divided mirror surfaces depending on filling ratio of PS wall. The thermal conductance ratio of divided mirror surface is bounded in the upper (lower) side when the gap is filled with PS (air).

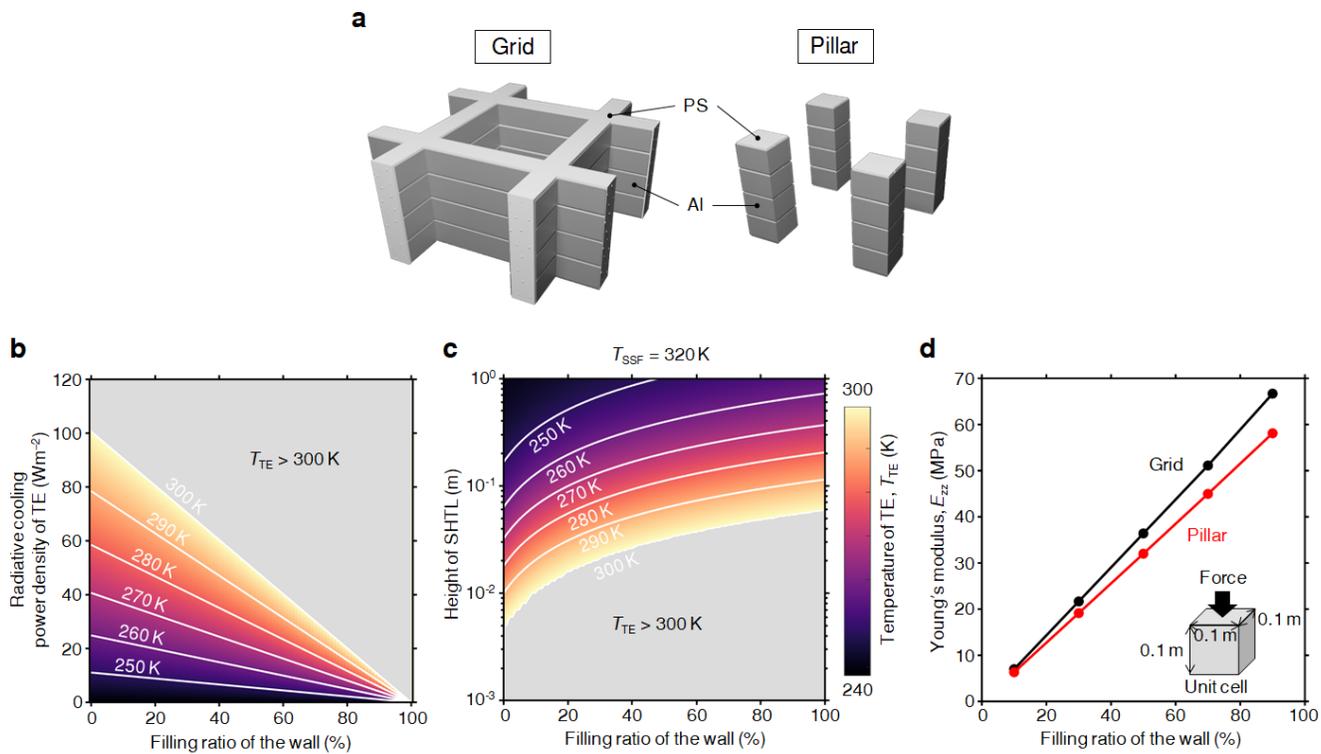

**Fig. S5 Thermal and physical properties according to SHTL design.** (**a**) Schematic for grid SHTL and pillar SHTL. (**b**) Radiative cooling power density of TE depending on the filling ratio of SHTL wall. (**c**) The steady-state temperature of TE depending on the filling ratio of the wall and the height of SHTL. (**d**) Young's modulus of SHTL, calculated by COMSOL simulation.

**Note S6: $h_{c,SHTL}$ extraction based on experimental data**

In this note, we present a method to extract non-radiative heat coefficient of designed SHTL. Usually, the non-radiative heat flows from hot SSF to cold TE but in finite-sized system, there can be heat dissipation through the sidewalls of SHTL. To consider the parasitic heat flow, one can apply 2D heat transfer model (Fig. S6a) that follows the energy conservation law as

$$\dot{Q}(z) = \dot{Q}(z + dz) + \dot{Q}_{parasitic}(z)$$
$$\approx \dot{Q}(z) + \frac{\partial \dot{Q}}{\partial z}\bigg|_z dz + \dot{Q}_{parasitic}(z) \quad (S6\text{-}1)$$

where $\dot{Q}(z) = -h_{c,SHTL} H A \frac{dT}{dz}$ is the heat transfer rate in which $H$ is the height of SHTL and $A$ is the area of SHTL in x-y plane, and $\dot{Q}_{parasitic}(z) = h_{c,parasitic} L (T(z) - T_{parasitic}) dz$ is the parasitic heat transfer rate in which $h_{c,parasitic}$ is the parasitic non-radiative heat coefficient, $L$ is the circumference length of the sidewall, and $T_{parasitic}$ is the parasitic temperature. Using these formulations, equation (S6-1) can be arranged as

$$\frac{d^2 T(z)}{dz^2} - \frac{h_{c,parasitic}}{h_{c,SHTL}} \frac{L}{AH} (T(z) - T_{parasitic}) = 0. \quad (S6\text{-}2)$$

Then, the solution of Eq. (S6-2) can be represented as

$$\therefore T(z) = a \cdot \exp(-bz) + c \quad (S6\text{-}3)$$

where the coefficients are $a = \frac{\dot{Q}(0)}{bAHh_{c,SHTL}}$, $b = \sqrt{\frac{h_{c,parasitic}}{h_{c,SHTL}} \frac{L}{AH}}$, and $c = T_{parasitic}$. There are two unknown variables ($h_{c,SHTL}$ and $h_{c,parasitic}$) in Eq. (S6-3), and the remaining parameters can be obtained from the experiments as follows: $\dot{Q}(0)$ = 0.24 W, $A$ = 1.45×10$^{-3}$ m², $H$ = 0.06, 0.08, and 0.1 m, $L$ = 0.12 m, and $T_{parasitic}$ is the temperature of baseline. Meanwhile, we can also solve the circuit model in Fig. S6b, which is equivalent to the above heat transfer model, such that $R_{SHTL} = dz H^{-1} A^{-1} h_{c,SHTL}^{-1}$, $R_{parasitic} = dz^{-1} L^{-1} h_{c,SHTL}^{-1}$.

Based on this model, one can fit the measured temperature of SHTL along the depth with Eq. (S6-3) to obtain $h_{c,SHTL}$. As a heat source, we applied a self-regulating resistive heater (SmartHeat SLT heater, MINCO Inc.) on SSF side, which is covered by wide and thick VIPs to prevent heat flow into the any direction except the bottom side. In addition, aluminum shields were applied to both sides of SSF and TE to inhibit radiative heat exchange through SHTL. Figures S6(b-d) show that measured and fitted temperatures agree well for different heights of SHTL. From the extracted coefficients, one can calculate $h_{c,SHTL}$ and $h_{c,parasitic}$, as shown in Fig. S6(f) and Table S2. The results for $h_{c,SHTL}$ show that TE side is highly insulated from SSF side, and the insulation level becomes stronger as the height of SHTL increases. Particularly, for 10 cm height of SHTL, $h_{c,SHTL}$ is almost 1 $Wm^{-2}K^{-1}$, which accords with the insulation of 30 cm thick polystyrene foam that allows only 40 $Wm^{-2}$ heat flow for 40 °C temperature difference.

On the other hand, $h_{c,parasitic}$ may be quite higher than $h_{c,SHTL}$, so we further examine how the parasitic interaction affects the cooling performance of RINE system. When SSF and TE non-radiatively interact with the parasitic site in a temperature $T_{parasitic}$, the power equations in the manuscript (Eqs. (1, 2)) can be revised as

$$P_{SSF} = W_{SSF} + \phi_{SSF,atm} + \phi_{SSF,SHTL} + \phi_{SSF,parasitic} \qquad (S6\text{-}4)$$

$$P_{TE} = W_{TE} + \phi_{TE,SHTL} + \phi_{TE,parasitic} \qquad (S6\text{-}5)$$

where $\phi_{SSF,parasite} = h_{c,parasite,eff}(T_{SSF} - T_{parasite})$ and $\phi_{TE,parasite} = h_{c,parasite,eff}(T_{TE} - T_{parasite})$ are the parasitic non-radiative heat conductance of SSF and TE, in which $h_{c,parasite,eff}$ is an effective parasitic non-radiative heat coefficient, relevant with $R_{parasitic,eff}$ in Fig S6b. And, the other terms are the same with that of previous equations. By solving Eqs. (S6-4, S6-5), one can obtain the steady-state temperature of SSF and TE, as shown in Fig. S7; in this case, $T_{parasitic} = T_{amb}$. The results indicate that, as $h_{c,parasitic}$ increases, the temperatures of SSF and TE converge to $T_{parasitic}$, thus SSF is heated less and TE is cooled less. The decreased temperature gap between SSF and TE may reduce non-radiative heat transfer through SHTL but the parasitic interaction essentially degrades sub-ambient cooling performance of TE even though $h_{c,SHTL}$ is small. Therefore, to improve cooling performance of TE, $h_{c,parasitic}$ should be lowered. Noteworthy is that the concern for parasitic interaction can be relieved in large-scale applications, such as façade of building, and rooftop of houses, that can satisfy the thermal periodic condition, thus $h_{c,parasitic}$ can be significantly smaller than on a finite-sized system.

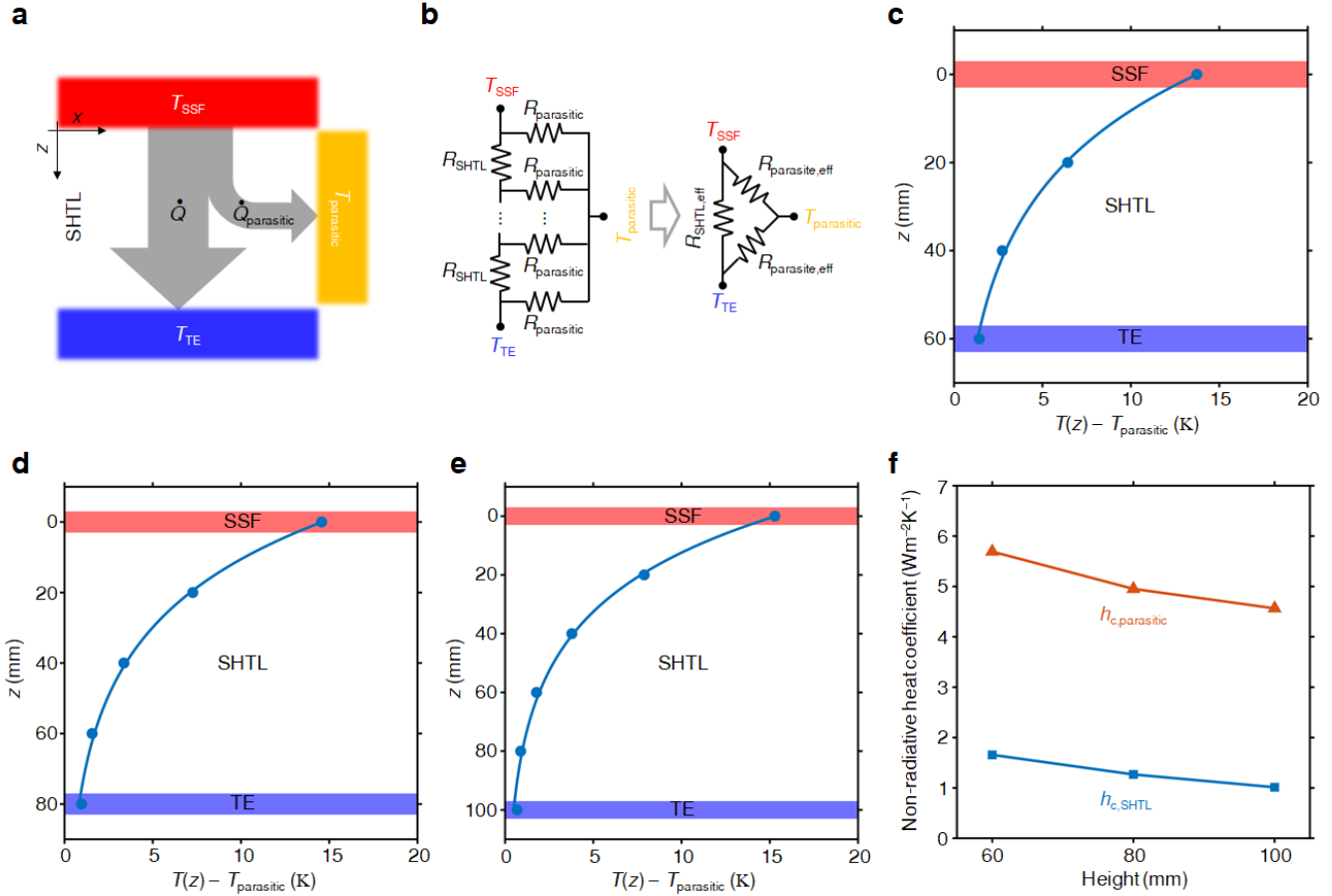

**Fig. S6 Temperature of SHTL along the depth and the extracted $h_{c,SHTL}$.** (**a**) Schematic and (**b**) equivalent circuit model of SHTL for finite-sided system. Experimentally measured temperatures (dots) and fitted lines by a heat transfer model along the depth of SHTL for the height of (**c**) 60 mm, (**d**) 80 mm, and (**e**) 100 mm. (**f**) Extracted $h_{c,SHTL}$ and $h_{c,parasitic}$.

| SHTL height | a (K) | b (m$^{-1}$) | c (K) | $h_{c,SHTL}$ (Wm$^{-2}$K$^{-1}$) | $h_{c,parasitic}$ (Wm$^{-2}$K$^{-1}$) |
|---|---|---|---|---|---|
| 60 mm | 13.8 | 38.8 | 300 | 1.66 | 5.69 |
| 80 mm | 14.6 | 35.9 | 299 | 1.27 | 4.95 |
| 100 mm | 15.3 | 34.4 | 298 | 1.01 | 4.57 |

**Table S2. Extracted parameters based on 2d heat transfer model.**

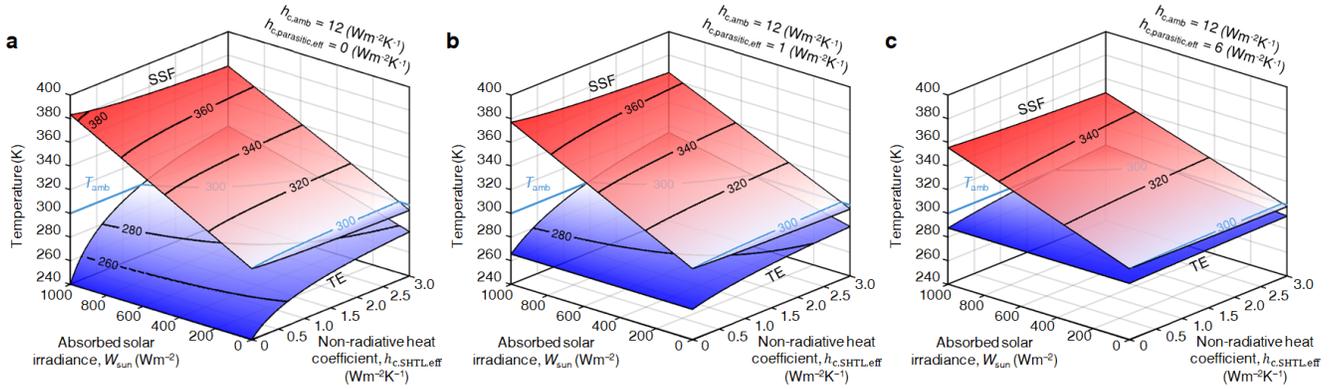

**Fig. S7 The steady-state temperature of finite-sized RINE system.** The spectral shapes of SSF and TE are the same with that described in the manuscript, which respectively accord with SSF type A and broadband TE in Fig. S2. In these plots, $h_{c,amb}$ is constant as 12 Wm$^{-2}$K$^{-1}$ while $h_{c,parasitic,eff}$ are **(a)** 0 Wm$^{-2}$K$^{-1}$, **(b)** 1 Wm$^{-2}$K$^{-1}$, and **(c)** 6 Wm$^{-2}$K$^{-1}$.

**Note S7: Cooling performance of designed RINE system**

As a validation of the experimentally designed samples, we calculate the cooling performance of RINE system that follows the spectral properties of SSF-red and TE in Note S10. Considering the spectral filtering of SSF, the net radiative cooling power density of TE is calculated for different $W_{sun}$, as shown in Fig. S8(a, b). As comparative systems, we consider the optimal RINE system (composed of SSF type A and broadband emitter in Fig. S2), the ideal emitter and two isothermal emitters whose emission band are selective in 8–13 μm and broadband over 4 μm[3]. For $W_{sun} = 0$, the net cooling power density of designed RINE system is over 85 Wm$^{-2}$ at $T_{amb}$, and even exceeds that of broadband emitter at low temperature. More importantly, for $W_{sun} = 719$ Wm$^{-2}$, the designed RINE system shows better cooling power compared to both of the isothermal emitters; in this case, the specific value of $W_{sun}$ (719 Wm$^{-2}$) is set for SSF-red at noontime of mid-latitude summer. Based on these results, we calculate the steady-state temperature depending on the effective non-radiative heat coefficient $h_{c,eff}$ (Fig. S8(c, d)) that can be defined as $h_{c,eff} = h_{c,amb}$ in an isothermal system, and as $h_{c,eff} = (h_{c,amb}^{-1} + h_{c,SHTL}^{-1})^{-1}$ in RINE system. For easy comparison, we set $h_{c,amb} = 12$ Wm$^{-2}$K$^{-1}$ for RINE system, indicating $h_{c,eff} = (12^{-1} + h_{c,SHTL}^{-1})^{-1}$. For $W_{sun} = 0$, TE of designed RINE system is cooled below $T_{amb}$, similar to the isothermal broadband emitter. Also, for $W_{sun} = 719$ Wm$^{-2}$, the steady-state temperature of designed TE is much lower than that of the other isothermal emitters, and even becomes lower than $T_{amb}$ when $h_{c,eff} < 1.3$ Wm$^{-2}$K$^{-1}$ that corresponds to $h_{c,SHTL} < 1.4$ Wm$^{-2}$K$^{-1}$. We note that extracted $h_{c,SHTL}$ in Note S6 fulfills such criteria for sub-ambient cooling of TE. Therefore, the calculations in this note show why the designed RINE system experimentally implemented sub-ambient cooling with vivid color representation.

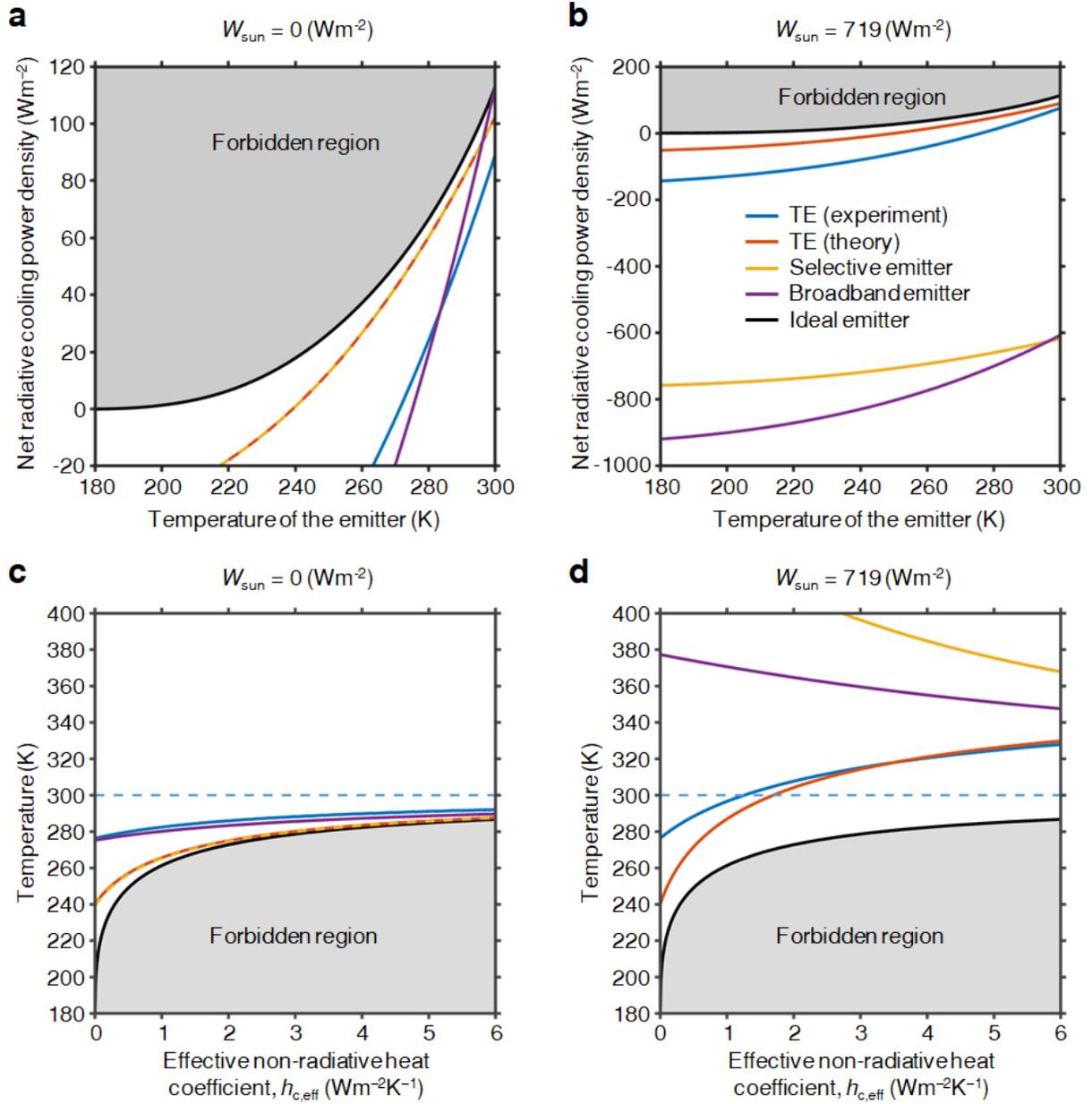

**Fig. S8 Radiative cooling performance of designed RINE system compared with the optimal systems.** Net radiative cooling power densities of RINE systems and isothermal emitters for (**a**) $W_{sun} = 0$ Wm$^{-2}$ and (**b**) $W_{sun} = 719$ Wm$^{-2}$. In RINE systems, $W_{sun}$ indicates the absorbed solar irradiances by SSF and the net cooling power density is for TE under SSF. The steady-state temperature of TEs and isothermal emitters for (**c**) $W_{sun} = 0$ Wm$^{-2}$ and (**d**) $W_{sun} = 719$ Wm$^{-2}$, along the effective non-radiative heat coefficient ($h_{c,eff}$); $h_{c,eff} = (12^{-1} + h_{c,SHTL}^{-1})^{-1}$ in RINE system, and $h_{c,eff} = h_{c,amb}$ in the isothermal system.

**Note S8: Spectral property of BB filter**

To analyze the experimental results for the reference sample described in the manuscript, we show the spectral absorptance (or emittance) of BB filter. As shown in Fig. S9, BB filter presents extremely high absorptance in the solar range and high emittance in the mid-IR range, just like a blackbody. These results note that TE under BB filter cannot radiatively interact with the outer surroundings but only with BB filter. Such strong radiative coupling between TE and top filter can impede the sub-ambient cooling of TE, mainly when SSF substantially absorbs the sunlight like BB filter, as demonstrated in Note S2.

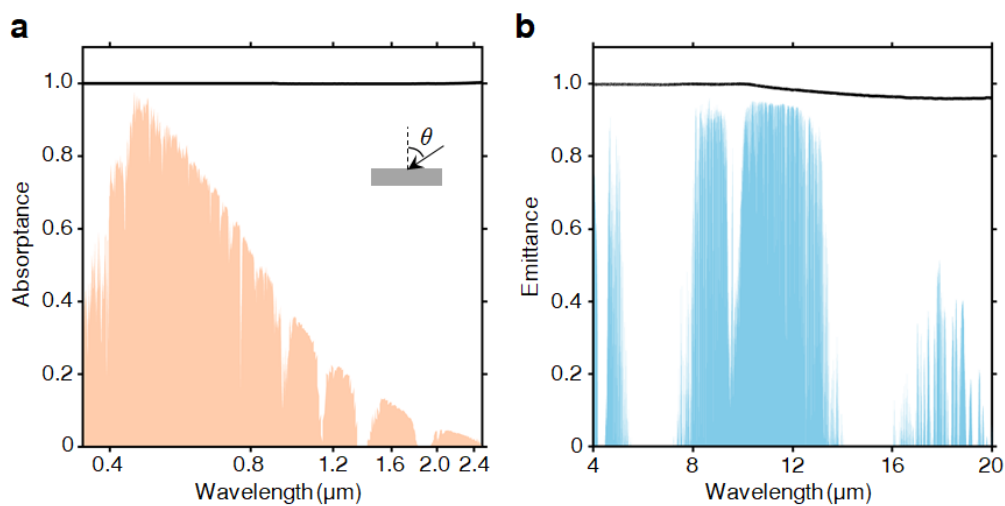

**Fig. S9 Spectral properties of BB filter.** (**a**) Absorptance in the solar range for 8° incidence angle. (**b**) Emittance in the mid-IR range for 30° incidence angle.

**Note S9: Climate conditions in measurement day**

For clarity, we present additional environmental data on the measurement day in Fig. S10, obtained from the Korea meteorological data center. Figure S10a illustrates the solar elevation angle at the measurement place (KAIST, 36.4 °N, 127.4 °E), which presents the maximum angle of 64.2° around 13:00 while the maximum of solar irradiance was recorded slightly earlier on a horizontal surface as 743 Wm$^{-2}$. The humidity was measured to be high near sunrise (maximum of 87%) and low near sunset (minimum of 36%), as shown in Fig. S10b. The wind was strong at over 3 ms$^{-1}$ throughout the day after 10:00 (Fig. S10c).

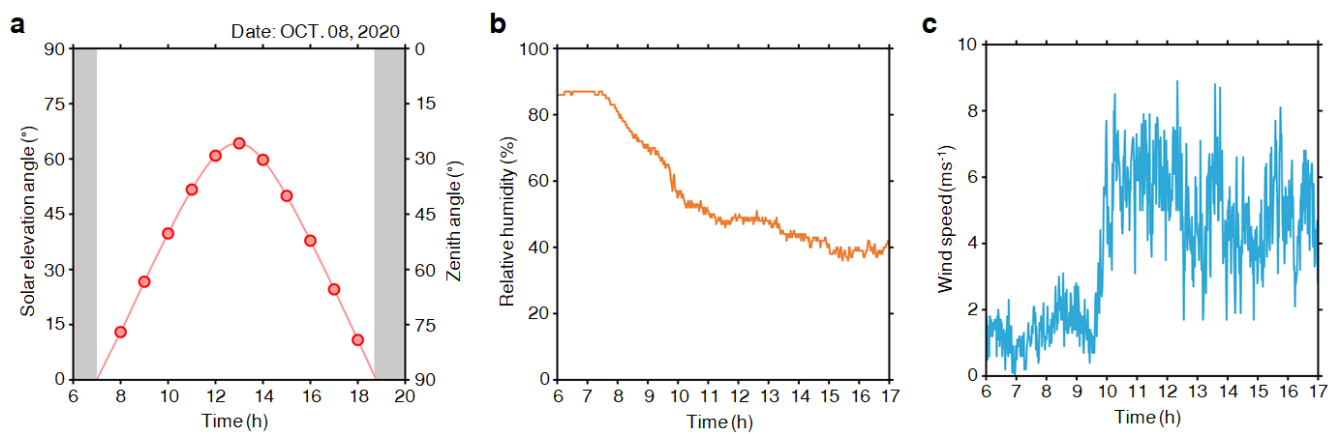

**Fig. S10 Climate conditions in measurement day.** (**a**) Solar elevation angle. (**b**) Relative humidity of the ambient air. (**c**) Wind speed.

**Note S10: Spectral angular properties of SSF and TE**

In this study, the angular character is essential for angle-tolerant color expression of SSF and effective radiative cooling of TE. So, we present the spectral angular reflectance of SSF-black, SSF-red, and TE in Fig. S11. The results for SSF-black and SSF-red show that their spectral reflectance in the solar range is robust to variation of viewing angle, which indicates consistent color expression of SSFs (Fig. S11(a, b)). Meanwhile, in the mid-IR range, directional properties can be beneficial to lower the temperature of TE further below $T_{amb}$ as disclosed in a previous work, which has revealed that radiative interaction at high zenith angle, particularly over 60°, degrades cooling performance due to the anisotropy of atmospheric transparency[3]. Figure S11(c, d) illustrates that the spectral selective properties of SSFs, aimed for anti-reflection at 8−13 μm wavelengths, fade out, and their reflectance rise at overall wavelengths as the incidence angle increases. These results indicate that SSFs hinder TE from exchanging radiative heat with surroundings through high zenith angles, which can be advantageous for sub-ambient cooling of TE. Moreover, the designed TE itself presents angle-dependent reflectance that accords with directional emission toward the zenith direction (Fig. S11f); due to zero transmittance of TE, the rest of reflectance is the emittance. For easy awareness of angular properties described above, we illustrate the anti-reflection of SSFs and the emittance of TE in Fig. S11f, by spectrally averaging them in 8–13 μm wavelengths.

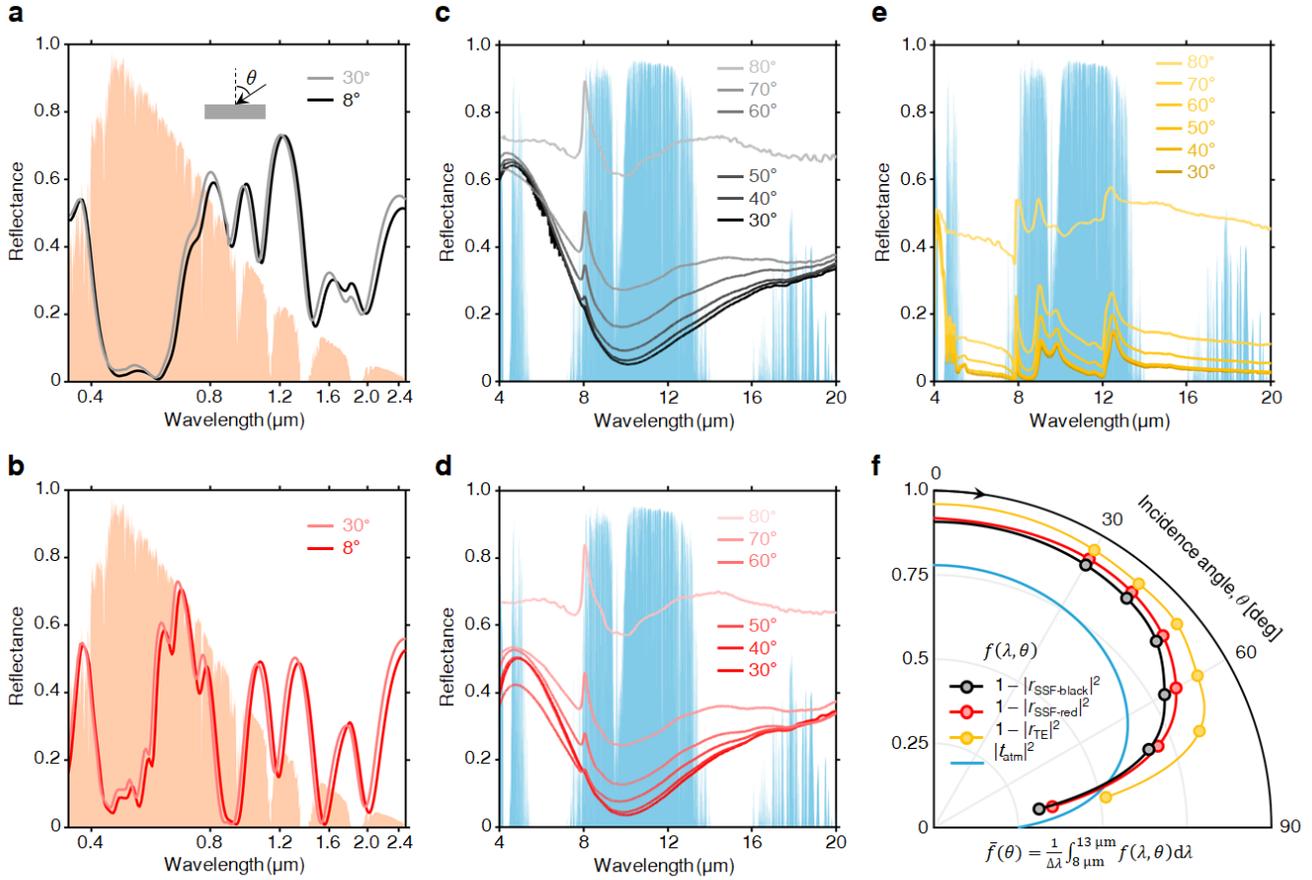

**Fig. S11 Spectral angular properties of SSF and TE.** Spectral reflectance of (**a**, **c**) SSF-black, (**b**, **d**) SSF-red, and (**e**) TE for various incidence angles. (**f**) Anti-reflectance (SSF-black and SSF-red), emittance (TE) and atmospheric transmittance, which are spectrally averaged from 8 to 13 μm, depending on incidence angle.

**Supplemental references**